\documentclass[twocolumn]{aastex63}

\usepackage{amsmath}
\usepackage{amssymb}
\usepackage{rotating}

\usepackage[T1]{fontenc}
\usepackage[utf8]{inputenc}

\mathchardef\mhyphen="2D 

\newcommand{\ra}[4]{${#1}^{\rm h}{#2}^{\rm m}{#3}\fs{#4}$}
\newcommand{\dec}[4]{${#1}\arcdeg{#2}\arcmin{#3}\farcs{#4}$}
\newcommand{\rashort}[3]{${#1}^{\rm h}{#2}^{\rm m}{#3}^{\rm s}$}
\newcommand{\decshort}[3]{${#1}\arcdeg{#2}\arcmin{#3}\arcsec$}

\newcommand\tE{t_{\rm E}}

\newcommand\piEN{\pi_{\textrm{E},N}}
\newcommand\piEE{\pi_{\textrm{E},E}}

\newcommand\muvec{\boldsymbol{\mu}}

\newcommand\thetaE{\theta_{\rm E}}

\newcommand\Ds{D_{\rm s}}
\newcommand\Dl{D_{\rm l}}

\received{}
\revised{}
\accepted{}

\shorttitle{Microlensing Optical Depth toward the SMC}
\shortauthors{P. Mr\'oz et al.}

\defcitealias{mroz2024b}{Paper~1}
\defcitealias{mroz2024a}{Paper~2}
\defcitealias{mroz2024c}{Paper~3}

\begin{document}

\title{Microlensing optical depth, event rate, and limits on compact objects in dark matter based on 20~years of OGLE observations of the Small Magellanic Cloud}

\correspondingauthor{Przemek Mr\'oz}
\email{pmroz@astrouw.edu.pl}

\author[0000-0001-7016-1692]{Przemek Mr\'oz}
\affil{Astronomical Observatory, University of Warsaw, Al. Ujazdowskie 4, 00-478 Warszawa, Poland}

\author[0000-0001-5207-5619]{Andrzej Udalski}
\affil{Astronomical Observatory, University of Warsaw, Al. Ujazdowskie 4, 00-478 Warszawa, Poland}

\author[0000-0002-0548-8995]{Micha\l{} K. Szyma\'nski}
\affil{Astronomical Observatory, University of Warsaw, Al. Ujazdowskie 4, 00-478 Warszawa, Poland}

\author[0000-0002-7777-0842]{Igor Soszy\'nski}
\affil{Astronomical Observatory, University of Warsaw, Al. Ujazdowskie 4, 00-478 Warszawa, Poland}

\author[0000-0002-2339-5899]{Pawe\l{} Pietrukowicz}
\affil{Astronomical Observatory, University of Warsaw, Al. Ujazdowskie 4, 00-478 Warszawa, Poland}

\author[0000-0003-4084-880X]{Szymon Koz\l{}owski}
\affil{Astronomical Observatory, University of Warsaw, Al. Ujazdowskie 4, 00-478 Warszawa, Poland}

\author[0000-0002-9245-6368]{Rados\l{}aw Poleski}
\affil{Astronomical Observatory, University of Warsaw, Al. Ujazdowskie 4, 00-478 Warszawa, Poland}

\author[0000-0002-2335-1730]{Jan Skowron}
\affil{Astronomical Observatory, University of Warsaw, Al. Ujazdowskie 4, 00-478 Warszawa, Poland}

\author[0000-0001-9439-604X]{Dorota Skowron}
\affil{Astronomical Observatory, University of Warsaw, Al. Ujazdowskie 4, 00-478 Warszawa, Poland}

\author[0000-0001-6364-408X]{Krzysztof Ulaczyk}
\affil{Department of Physics, University of Warwick, Coventry CV4 7 AL, UK}
\affil{Astronomical Observatory, University of Warsaw, Al. Ujazdowskie 4, 00-478 Warszawa, Poland}

\author[0000-0002-1650-1518]{Mariusz Gromadzki}
\affil{Astronomical Observatory, University of Warsaw, Al. Ujazdowskie 4, 00-478 Warszawa, Poland}

\author[0000-0002-9326-9329]{Krzysztof Rybicki}
\affil{Department of Particle Physics and Astrophysics, Weizmann Institute of Science, Rehovot 76100, Israel}
\affil{Astronomical Observatory, University of Warsaw, Al. Ujazdowskie 4, 00-478 Warszawa, Poland}

\author[0000-0002-6212-7221]{Patryk Iwanek}
\affil{Astronomical Observatory, University of Warsaw, Al. Ujazdowskie 4, 00-478 Warszawa, Poland}

\author[0000-0002-3051-274X]{Marcin Wrona}
\affil{Department of Astrophysics and Planetary Science, Villanova University, 800 East Lancaster Avenue, Villanova, PA 19085, USA}
\affil{Astronomical Observatory, University of Warsaw, Al. Ujazdowskie 4, 00-478 Warszawa, Poland}

\author[0000-0002-3218-2684]{Milena Ratajczak}
\affil{Astronomical Observatory, University of Warsaw, Al. Ujazdowskie 4, 00-478 Warszawa, Poland}

\begin{abstract}
Some previous studies have suggested that massive and intermediate-mass primordial black holes (PBHs) could comprise a substantial fraction of dark matter in the Universe. Such black holes, if they existed in the Milky Way halo, would give rise to long-duration microlensing events that may potentially last for years. However, earlier searches were not sufficiently sensitive to detect such events. Here, we present the results of searches for long-timescale gravitational microlensing events toward the Small Magellanic Cloud (SMC) using nearly 20\,yr of photometric observations collected by the Optical Gravitational Lensing Experiment (OGLE) from 2001--2020. We found six events, three of which are new discoveries. We use a sample of five events to measure the microlensing optical depth toward the SMC $\tau = (0.32 \pm 0.18) \times 10^{-7}$ and the event rate $\Gamma = (1.18 \pm 0.57) \times 10^{-7}\,\mathrm{yr}^{-1}\,\mathrm{star}^{-1}$. The properties of the detected events are consistent with lenses originating from known stellar populations within the SMC or in the Milky Way disk. No events with timescales longer than 1\,yr were detected, which provides competitive limits on the fraction of massive compact objects, including PBHs, in the Milky Way dark matter halo. Together with the earlier OGLE studies of microlensing events in the direction of the Large Magellanic Cloud, these observations rule out PBHs and other compact objects with masses ranging from $10^{-8}$--$10^3\,M_{\odot}$ as dominant components of dark matter.
\end{abstract}

\keywords{Gravitational microlensing (672); Dark matter (353); Milky Way dark matter halo (1049); Small Magellanic Cloud (1468); Primordial black holes (1292); Intermediate-mass black holes (816)}

\section{Introduction}

The hypothesis that primordial black holes (PBHs) may account for a substantial part of dark matter has gained some popularity in recent years, especially after the discoveries of mergers of unexpectedly massive black holes by gravitational-wave detectors LIGO and Virgo \citep[e.g.,][]{bird2016,sasaki2016,clesse2017}. PBHs are hypothesized to have formed in the very early Universe by the collapse of large density perturbations \citep[e.g.,][]{zeldovich1967,hawking1971,carr1974,chapline1975}. Because the formation of PBHs is thought to have occurred before the Big Bang nucleosynthesis, PBHs are nonbaryonic, and their current abundance is not constrained by primordial elemental abundances \citep[e.g.,][and references therein]{navas2024}. Although PBHs cannot be directly observed, their existence would manifest in a number of phenomena, including (but not limited to) gravitational microlensing of background stars, emission of Hawking radiation, production of gravitational waves, accretion of gas, impact on the dynamics of dwarf galaxies and wide binary star systems, or cosmological structure formation (see the reviews by \citealt{green2021}; \citealt{carr2021}; \citealt{carr2024}).

Despite this rich phenomenology, searches for PBH signatures have not provided compelling arguments for the PBH dark matter hypothesis. Instead, these investigations have led to numerous constraints on the amount of the PBH dark matter in the Universe as a function of the PBH mass. These constraints are often calculated under the assumption of a smooth distribution of dark matter in galactic halos. However, numerical simulations of galaxy formation indicate that dark matter halos may contain numerous substructures, with characteristic sizes depending on the specific dark matter model \citep[e.g.,][]{bechtol2022}. They may be detected, for example, using observations of strongly lensed galaxies. \citet{vegetti2010,vegetti2012} reported the discovery of such substructures with masses of the order of $10^8\!-\!10^9\,M_{\odot}$ but it is unclear whether they are associated with lensing galaxies or some unrelated structures along the line of sight \citep[e.g.,][]{li2017,despali2018,sengul2022}.

Some of the strongest constraints on the abundance of PBHs with masses from $10^{-8}\,M_{\odot}$ to $10^3\,M_{\odot}$ come from searches for gravitational microlensing events in the long-term photometric observations of stars in the Magellanic Clouds, as first proposed by \citet{paczynski1986}. The long-term observations of the Large Magellanic Cloud (LMC) by the Optical Gravitational Lensing Experiment \citep[OGLE;][hereafter \citetalias{mroz2024b} and \citetalias{mroz2024a}, respectively]{mroz2024b,mroz2024a} demonstrate that less than 1\% of dark matter may be composed of compact objects with masses from $1.8\times 10^{-4}$ to $6.3\,M_{\odot}$. High-cadence OGLE observations of the LMC and the Small Magellanic Cloud (SMC) provide strong limits on the abundance of planetary-mass PBHs \citep[][hereafter, \citetalias{mroz2024c}]{mroz2024c}. These limits supersede the earlier constraints derived from searches for microlensing events toward the LMC by MACHO \citep{alcock1997c,alcock2000b}, EROS \citep{tisserand2007,moniez2022}, \mbox{OGLE-II} \citep{wyrzykowski2009}, and \mbox{OGLE-III} \citep{wyrzykowski2011} experiments by more than an order of magnitude. However, to probe possible dark matter substructures in the Milky Way halo, it is important to search for gravitational microlensing events toward different sky directions.

One possible target is the Andromeda galaxy (M31). The searches for massive compact halo objects in the direction of M31 by \citet{calchi_novati2005} and \citet{de_jong2006} yielded contradictory results due to the difficulty in distinguishing self-lensing events (in which stars from M31 are lensing stars in the same galaxy) from those by PBHs (see also \citealt{calchi_novati2010}). On the other hand, the high-cadence observations of M31 by the Subaru Hyper Suprime-Cam (HSC) survey \citep{niikura2019} provided strong limits on the abundance of moon- to planetary-mass PBHs ($10^{-10}\!-\!10^{-6}\,M_{\odot}$) in dark matter. 

The SMC was also a target of microlensing surveys. The MACHO project, which operated from 1992 to 1999, reported the discovery of two microlensing events toward this galaxy \citep{alcock1997_smc,alcock1999}. The EROS-2 survey (1996--2003) detected only one event. Only one event in the SMC was found in the \mbox{OGLE-II} data (1996--2000) by \citet{wyrzyk3}, and three events were found in the \mbox{OGLE-III} data (2001--2009) by \citet{wyrzyk1}. These small numbers were consistent with the expected number of self-lensing events \citep{sahu1994,calchi_novati2013,mroz_poleski2018}, in which both the microlensing source and lens are located in the same galaxy, and events by nearby lenses in the Milky Way disk \citep{gould1994ApJ}. No PBHs were required to explain these events.

Previous SMC surveys were not sensitive to very long-duration microlensing events (with timescales of several years) that are expected from massive and intermediate-mass black holes, similar to those detected in large numbers by the LIGO and Virgo observatories. In this study, we combined the data collected during the third and fourth phases of the OGLE project to create nearly 20\,yr long, uniform light curves for about 7.4 million stars in the SMC. In addition, 6.2~million stars were observed by \mbox{OGLE-IV} only. This paper presents the results of our search for the very long-duration microlensing events in this data set. 

The data set analyzed in this study is described in Section~\ref{sec:data}. The selection criteria for microlensing events are detailed in Section~\ref{sec:search}, while Section~\ref{sec:events} presents the properties of the events that we discovered. Sections \ref{sec:stars}, \ref{sec:eff}, and~\ref{sec:tau} outline the calculations of the number of source stars, detection efficiencies, and the microlensing optical depth in the direction of the SMC, respectively. In Section~\ref{sec:sl}, we demonstrate that the detected events can be explained by the known stellar populations in both the Milky Way and the SMC. Finally, in Sections~\ref{sec:limits} and \ref{sec:dis}, we calculate the limits on the PBH abundance in dark matter and discuss our results.

\begin{figure*}
\centering
\includegraphics[width=\textwidth]{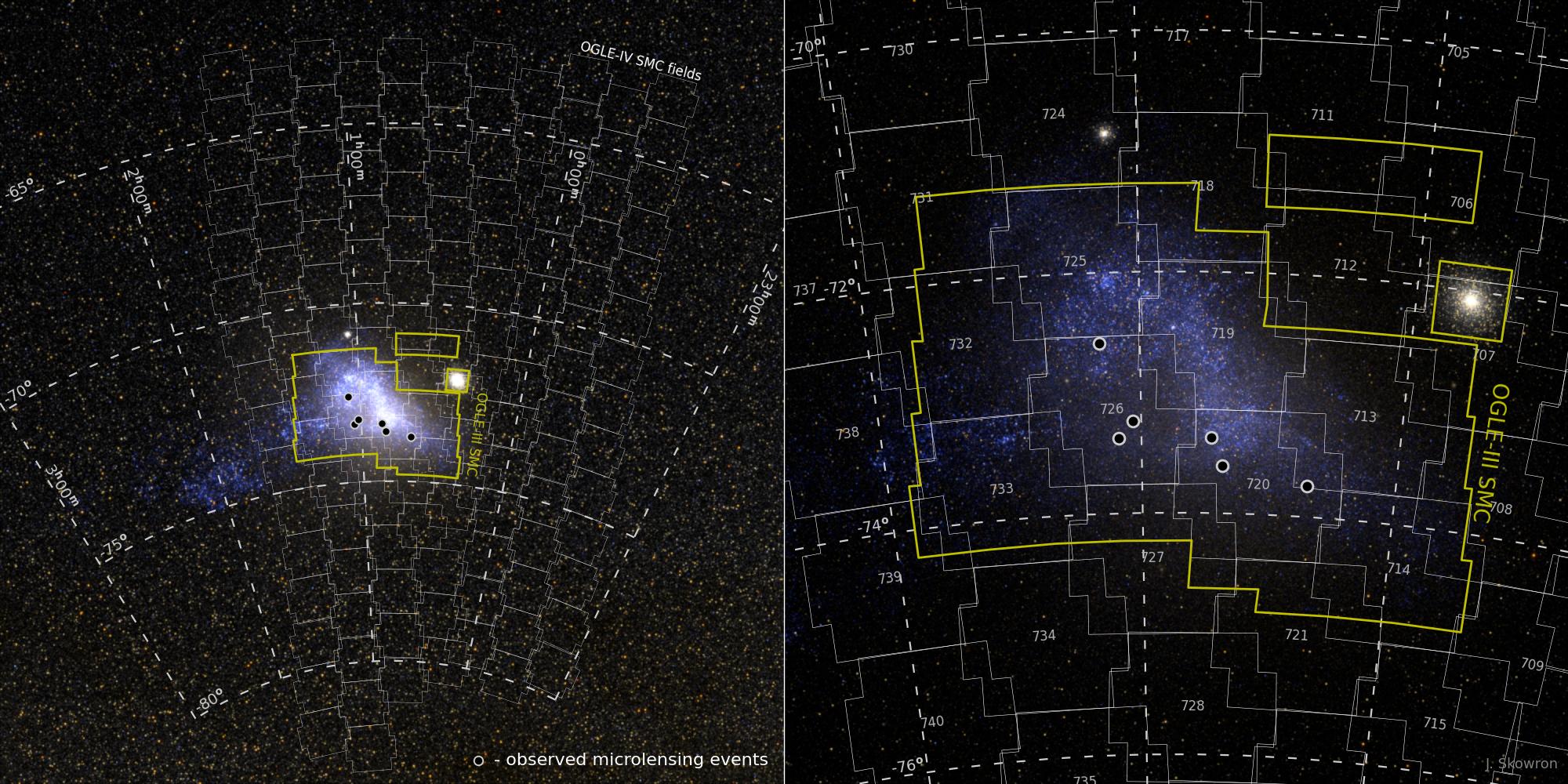}
\caption{\mbox{OGLE-IV} fields toward the Small Magellanic Cloud (SMC) are shown with white polygons. The numbers in the center of each polygon in the right panel indicate the field number. The yellow lines mark the regions that were observed during the \mbox{OGLE-III} phase. Detected microlensing events are marked with black circles. The background image of the SMC was generated with \textsc{bsrender} written by Kevin Loch, using data from the ESA/Gaia database.}
\label{fig:fields}
\end{figure*}

\begin{deluxetable*}{lllrrrrr}
\centering
\tablecaption{OGLE-IV Fields toward the Small Magellanic Cloud\label{tab:fields}}
\tablehead{
\colhead{Field} & \colhead{R.A.} & \colhead{Decl.} & \colhead{$N_{\rm Epochs}$}& \colhead{$N_{\rm Epochs}$} & \colhead{$N_{\rm Stars}$} & \colhead{$N_{\rm Stars}$} & \colhead{$N_{\rm Stars}$}\\
\colhead{} & \colhead{(J2000)} & \colhead{(J2000)} & \colhead{(2001--2009)} & \colhead{(2010--2020)} & \colhead{(2001--2020)} & \colhead{(2010--2020)} & \colhead{(Total)}}
\startdata
SMC700 & \rashort{00}{15}{21} & \decshort{-69}{26}{37} & 0       &  679 &     0.0 & 61.8 &  61.8 \\
SMC701 & \rashort{00}{14}{03} & \decshort{-70}{40}{51} & 0       &  677 &     0.0 & 67.4 &  67.4 \\
SMC702 & \rashort{00}{11}{02} & \decshort{-71}{54}{43} & 0       &  674 &     0.0 & 79.4 &  79.4 \\
SMC703 & \rashort{00}{05}{33} & \decshort{-73}{08}{19} & 0       &  673 &     0.0 & 68.6 &  68.6 \\
SMC704 & \rashort{00}{28}{58} & \decshort{-68}{49}{55} & 0       &  667 &     0.0 & 59.0 &  59.0 \\
SMC705 & \rashort{00}{28}{34} & \decshort{-70}{03}{44} & 0       &  668 &     0.0 & 68.9 &  68.9 \\
SMC706 & \rashort{00}{26}{26} & \decshort{-71}{17}{37} & 0--1377 &  666 &    90.7 & 25.9 & 116.6 \\
SMC707 & \rashort{00}{22}{01} & \decshort{-72}{31}{19} & 0--1371 & 1090 &   336.0 & 40.6 & 376.7 \\
SMC708 & \rashort{00}{17}{23} & \decshort{-73}{45}{14} & 0--1353 &  658 &    89.7 & 84.2 & 173.9 \\
SMC709 & \rashort{00}{10}{13} & \decshort{-74}{59}{06} & 0--663  &  654 &     3.8 & 76.4 &  80.2 \\
\multicolumn{1}{c}{\dots} & \multicolumn{1}{c}{\dots} & \multicolumn{1}{c}{\dots} & \multicolumn{1}{c}{\dots} & \multicolumn{1}{c}{\dots} & \multicolumn{1}{c}{\dots} & \multicolumn{1}{c}{\dots} & \multicolumn{1}{c}{\dots} \\
Total & \multicolumn{1}{c}{\dots} & \multicolumn{1}{c}{\dots} & \multicolumn{1}{c}{\dots} & \multicolumn{1}{c}{\dots} & 7431.2 & 6180.1 & 13611.4 \\
\enddata
\tablecomments{The coordinates are given for the epoch J2000. $N_{\rm Epochs}$ is the number of $I$-band epochs, separately for \mbox{OGLE-III} (2001--2009) and \mbox{OGLE-IV} (2010--2020) phases of the survey. $N_{\rm Stars}$ is the number of stars (in thousands) detected in the reference image of a given field. This number is not equal to the number of microlensing source stars that enter the optical depth calculations. We separately provide the number of stars observed during both \mbox{OGLE-III} and \mbox{OGLE-IV} (that is, from 2001--2020) and during \mbox{OGLE-IV} only (2010--2020). Columns (6) and (7) may not sum to column (8) due to rounding. (This table is available in its entirety in machine-readable form.)}
\end{deluxetable*}

\section{Data}
\label{sec:data}

\subsection{OGLE-IV Data}

The photometric data that are analyzed in this paper were collected during the fourth phase of the OGLE survey (\mbox{OGLE-IV}; \citealt{udalski2015}) between June 29, 2010 ($\mathrm{JD}=2,455,377$) and March 15, 2020 ($\mathrm{JD}=2,458,924$), when OGLE operations were halted because of the start of the \mbox{COVID-19} pandemic. The survey uses a dedicated 1.3\,m Warsaw Telescope located at Las Campanas Observatory, Chile. The telescope is equipped with a mosaic CCD camera consisting of 32 detectors, each with a resolution of $2048 \times 4102$ pixels. With a pixel scale of $0.26$\,arcsec\,pixel$^{-1}$, the \mbox{OGLE-IV} camera covers the field of view of approximately 1.4\,deg$^2$. Further details about the instrument can be found in \citet{udalski2015}.

The \mbox{OGLE-IV} project observes an area of approximately 190\,deg$^2$ surrounding the SMC. This region is divided into 135 fields, whose locations are shown in Figure~\ref{fig:fields}. The equatorial coordinates of the center of each field, along with the number of epochs and the number of detected stars, are provided in Table~\ref{tab:fields}. Most of the data were taken in the $I$-band filter, whose transmission curve closely resembles that of the standard Cousins photometric system. The number of $I$-band epochs varies from 151 to 1090, with a median count of 376. In addition, a smaller number of images were taken through the $V$-band filter (from 7 to 321 per field with a median of 22). The cadence of observations differs depending on the specific field. Fields covering the SMC center were observed at a 2--5\,day cadence, whereas the outer fields were observed less frequently, typically  every 4--6 days. The typical seeing ranged from $0.9$ to $2.0''$, with a median seeing of $1.4''$. Exposure times varied between 150 and 170\,s.

We employed the difference image analysis (DIA) technique \citep{tomaney1996,alard1998}, as implemented by \citet{wozniak2000}, to extract photometry. This method provides high-quality measurements, even in the most crowded stellar regions. For every analyzed field, a reference image is constructed by stacking several (from 2--10) low-seeing, low-background individual images. Photometry is performed on difference (subtracted) images, which are generated by subtracting the reference image from the incoming frames. Our photometric pipeline detects and flags stars that are identified in the subtracted images. In addition, forced photometry with fixed centroids is calculated for all stars detected in the reference images. In total, the analyzed fields contain about 13.6 million such stars. Please note that the number of stars that are detected on reference images is not equal to the number of microlensing source stars, as discussed in Section~\ref{sec:stars}.

\subsection{Combined OGLE-III/OGLE-IV Data}
\label{sec:data_o3_o4}

Because our main scientific goal was to search for long-duration microlensing events, we supplemented the \mbox{OGLE-IV} light curves with data collected during the third phase of the survey (\mbox{OGLE-III}; \citealt{udalski2003}), following the methodology detailed in \citetalias{mroz2024b}. The OGLE-III data were gathered from June 24, 2001 ($\mathrm{JD}=2,452,085$) to May 2, 2009 ($\mathrm{JD}=2,454,954$). This approach allowed us to extend the \mbox{OGLE-IV} light curves to nearly 20\,yr.

The \mbox{OGLE-III} project used the same Warsaw Telescope as \mbox{OGLE-IV} albeit with a smaller camera that contained eight $2048\!\times\!4096$ detectors. Observations were taken in the $I$-band filter, similar to that used by \mbox{OGLE-IV}, with exposure times of 180\,s. Because the reference images from \mbox{OGLE-III} and \mbox{OGLE-IV} were slightly different, we could not naively combine the light curves from both data sets. 
For instance, some close pairs of stars could be resolved in the \mbox{OGLE-III} reference images but not in \mbox{OGLE-IV} reference images, and vice versa. These differences could lead to systematic shifts between the two data sets. Fortunately, because the pixel scales of the \mbox{OGLE-III} and \mbox{OGLE-IV} cameras were identical, we were able to reduce the \mbox{OGLE-III} images using \mbox{OGLE-IV} reference images using the standard OGLE DIA pipeline by \citet{wozniak2000}. This procedure provided homogeneous light curves for all common stars. As demonstrated in \citetalias{mroz2024b}, the differences between the mean magnitudes from \mbox{OGLE-III} and \mbox{OGLE-IV} were consistent with zero (see Figure~2 in \citealt{mroz2024b}).

The \mbox{OGLE-III} project observed an area of 14\,deg$^2$ in the SMC, focusing on the central regions of this galaxy. This area is marked with yellow lines in Figure~\ref{fig:fields} and contains a total of 7.4~million stars (about 55\% of all SMC stars detected in \mbox{OGLE-IV} reference images). The \mbox{OGLE-III} fields were observed between 583 and 762 times. Table~\ref{tab:fields} provides more detailed statistics about the number of stars that have full 2001--2020 light curves and the total number of available epochs.

A smaller number of 2.1 million stars located in the central 2.4\,deg$^2$ of the SMC were observed during the second phase of the OGLE survey (\mbox{OGLE-II}; \citealt{udalski1997,wyrzyk3}) from 1997--2000. By combining this data set with the light curves from \mbox{OGLE-III} and \mbox{OGLE-IV}, we can extend the full data set to nearly 24,yr. However, the \mbox{OGLE-II} camera had a larger pixel size of $0.417''$ compared to its successors, which makes it nearly impossible to extract homogeneous light curves using the methodology developed in \citetalias{mroz2024b}. Therefore, we use the \mbox{OGLE-II} data, if available, only to vet the microlensing event candidates detected in the combined \mbox{OGLE-III}/\mbox{OGLE-IV} data set.

\subsection{Correction of Error Bars}
\label{sec:errors}

The uncertainties returned by the DIA pipeline do not accurately reflect the observed scatter in the light curves of constant stars \citep[e.g.,][]{skowron2016}. Therefore, they need to be rescaled using the formula $\delta m_{\rm new}=\sqrt{(\gamma \delta m_{\rm old})^2+\varepsilon^2}$, where the parameters $\gamma$ and $\varepsilon$ depend on the field and filter used. We followed the procedure described by \citet{skowron2016} to determine the values of these parameters, separately for the new \mbox{OGLE-III} and \mbox{OGLE-IV} light curves. Our results are reported in Table~\ref{tab:errorbars}. For the original \mbox{OGLE-IV} data, we found the median values of $\gamma=1.36$ and $\varepsilon=0.015$.

\begin{deluxetable}{ccccc}
\centering
\tablecaption{Error Bar Correction Coefficients for the $I$-band Data\label{tab:errorbars}}
\tablehead{
\colhead{Field} & \colhead{$\gamma_{\rm OGLE\mhyphen III}$} & \colhead{$\varepsilon_{\rm OGLE\mhyphen III}$} & \colhead{$\gamma_{\rm OGLE\mhyphen IV}$} & \colhead{$\varepsilon_{\rm OGLE\mhyphen IV}$}}
\startdata
SMC720.01 & 1.549 & 0.0082 & 1.660 & 0.0056 \\
SMC720.02 & 1.522 & 0.0085 & 1.500 & 0.0054 \\
SMC720.03 & 1.531 & 0.0086 & 1.517 & 0.0062 \\
SMC720.04 & 1.653 & 0.0077 & 1.574 & 0.0057 \\
SMC720.05 & 1.569 & 0.0076 & 1.506 & 0.0059 \\
SMC720.06 & 1.522 & 0.0078 & 1.454 & 0.0066 \\
SMC720.07 & 1.588 & 0.0079 & 1.520 & 0.0063 \\
SMC720.08 & 1.617 & 0.0080 & 1.536 & 0.0064 \\
SMC720.09 & 1.388 & 0.0069 & 1.458 & 0.0053 \\
SMC720.10 & 1.465 & 0.0084 & 1.529 & 0.0054 \\
\dots & \dots & \dots & \dots & \dots \\
\enddata
\tablecomments{(This table is available in its entirety in machine-readable form.)}
\end{deluxetable}

\section{Search for Microlensing Events}
\label{sec:search}

For our search for microlensing events in the SMC data, we employed a methodology very similar to that developed in \citetalias{mroz2024b}, with some minor modifications aimed at maximizing detection efficiency while minimizing contamination from non-microlensing light curves. For a detailed discussion of the event selection criteria, we refer readers to \citet{mroz2024b}. The list of all selection criteria is presented in Table~\ref{tab:cuts}.

We began with light curves of 13.6 million stars that were detected on \mbox{OGLE-IV} reference images. Of these, 7.4 million also had \mbox{OGLE-III} data available. We extracted their light curves as described in Section~\ref{sec:data_o3_o4}. The uncertainties returned by the DIA pipeline were corrected as discussed in Section~\ref{sec:errors}, and the magnitudes were subsequently converted into flux units.

In the first step, we searched for objects that underwent at least one significant brightening (called a ``bump'') in their light curves. Specifically, we required that at least five consecutive data points had to be brighter than $F_{\rm base} + 3 \sigma_{\rm base}$, where $F_{\rm base}$ is the mean flux, and $\sigma_{\rm base}$ is its standard deviation measured outside the magnified section of the light curve. Additionally, we required that the object be detected by the DIA pipeline on at least three subtracted images during the bump, its amplitude be at least 0.1 mag, and it show no significant variability beyond the main brightening. This initial filtering reduced the number of candidate events from 13.6 million to 25,973.

Next, we applied a series of selection cuts to eliminate apparent non-microlensing light curves. We first algorithmically removed the main bump and searched for any additional brightenings in the data. As microlensing events are expected to be non-repeating, we excluded objects that displayed more than one bump in their light curves. We also discarded ``blue bumper'' stars, which are defined as those located in the color--magnitude diagram (CMD) region where $(V-I)_0 \leq 0.5$ and $I_0 \leq 20.0$ (notably, the magnitude cut differs from that used in \citetalias{mroz2024b} to account for the larger SMC distance modulus). The de-reddened colors and magnitudes were calculated using the reddening map by \citet{skowron2021}. 

In addition to the cuts described in \citetalias{mroz2024b}, we excluded objects located within $10'$ of the globular cluster 47~Tuc, as their photometry was significantly affected by extreme crowding. We also decided to discard high proper-motion stars, which displayed significant trends in their light curves that were picked up by our algorithm as microlensing candidates. To this end, we cross-matched the list of selected candidates with the Gaia DR3 archive \citep{gaia2016,gaia_edr3} and removed any objects with proper motions greater than 10\,mas\,yr$^{-1}$. All of these criteria reduced the number of candidate events to 9614.

In the final step, we fitted a microlensing point-source point-lens (PSPL) model to the data and evaluated several goodness-of-the-fit statistics. In this model, the flux is given by the formula
\begin{equation}
F(t) = F_0 [1 + f_{\rm s} (A (t) - 1)],
\end{equation}
where $F_0$ is the baseline flux, $f_{\rm s}$ is the dimensionless blending parameter, and the magnification can be calculated using the equation
\begin{equation}
A(t) = \frac{u(t)^2+2}{u(t)\sqrt{u(t)^2+4}}.
\end{equation}
Here, $u(t)^2 = ((t-t_0)/t_{\rm E})^2 + u_0^2$, with $t_0$ being the time of the maximum magnification, $t_{\rm E}$ the Einstein radius crossing timescale, and $u_0$ the impact parameter in Einstein radius units.

\begin{table*}
\centering
\caption{Event Selection Criteria}
\label{tab:cuts}
\begin{tabular}{llr}
\hline \hline
& \multicolumn{1}{c}{Criteria} & \multicolumn{1}{c}{Number} \\
\hline
\textbf{Cut 0.} & \textbf{All stars in the databases} & 13,611,366 \\
\hline
\textbf{Cut 1.} & \textbf{Stars with at least one significant brightening in the light curve} \\
& At least five consecutive data points $3\sigma$ above the baseline flux ($n_{\rm bump} \geq 5$) & \\
& Object detected on at least three subtracted images ($n_{\rm DIA} \geq 3$) &\\
& The total ``significance'' of the bump ($\chi_{3+}=\sum_i(F_i-F_{\rm base})/\sigma_i \geq 32$) &\\
& Amplitude of the bump at least 0.1 mag ($\Delta m \geq 0.1$ mag) & \\
& Amplitude of the bump at least 0.4 mag if the bump is longer than 100 days & \\
& No significant variability outside the window ($\chi^2_{\rm out}/\mathrm{d.o.f.} \leq 2$) if the bump is shorter than 1000 days & 25,973 \\
\hline
\textbf{Cut 2.} & \textbf{Removing false positives} & \\
& Stars with multiple bumps in the data & \\
& ``Blue bumpers'' ($(V-I)_0 \leq 0.5$, $I_0 \leq 20.0$) & \\
& Stars in the vicinity of 47~Tuc & \\
& High proper-motion stars & 9614 \\
\hline
\textbf{Cut 3.} & \textbf{Microlensing PSPL model describes the light curve well} & \\
& Fit converged & \\
& $\chi^2/\mathrm{d.o.f.} \leq 2$ (all data points) & \\
& $\chi^2_{\tE}/\mathrm{d.o.f.} \leq 2$ (for $|t - t_0| < \tE$) & \\
& $\chi^2_{\rm bump}/\mathrm{d.o.f.} \leq 2$ (for data points within the bump) & \\
& Microlensing model significantly better than a straight line ($\chi^2_{\rm line}-\chi^2 \geq 250 \chi^2/\mathrm{dof}$) & \\
& $t_0$ within the time range covered by the data & \\
& Impact parameter is smaller than 1 ($u_0 \leq 1$) & \\ 
& Uncertainty on the Einstein timescale ($\sigma(\tE) / \tE \leq 1$) & \\
& Source star brighter than 22 mag ($I_{\rm s} \leq 22$~mag) & 5\\
\hline
\end{tabular}
\end{table*}

\begin{deluxetable*}{ccclccc}
\centering
\tablecaption{Detected Microlensing Events\label{tab:events}}
\tablehead{
\colhead{Event} & \colhead{R.A. (J2000)} & \colhead{Decl. (J2000)} & \colhead{OGLE-IV ID}& \colhead{$I_{\rm base}$ (mag)} & \colhead{$(V-I)_{\rm base}$ (mag)} & \colhead{$E(V-I)$ (mag)}}
\startdata
OGLE-SMC-02 & \ra{00}{40}{28}{12} & \dec{-73}{44}{46}{5} & SMC713.01.1243 & $18.427 \pm 0.010$ & $0.953 \pm 0.014$ & $0.035^{+0.026}_{-0.029}$\\
& & & SMC720.24.1511 & & & \\
OGLE-SMC-03 & \ra{01}{02}{37}{51} & \dec{-73}{22}{26}{0} & SMC726.12.12326                & $20.642 \pm 0.011$ & $0.067 \pm 0.016$ & $0.065^{+0.048}_{-0.053}$\\
OGLE-SMC-04 & \ra{01}{00}{59}{33} & \dec{-73}{14}{17}{1} & SMC726.13.27861                & $18.505 \pm 0.010$ & $0.971 \pm 0.014$ & $0.060^{+0.039}_{-0.048}$\\
OGLE-SMC-05 & \ra{01}{04}{29}{29} & \dec{-72}{35}{14}{9} & SMC726.28.35652                & $20.614 \pm 0.012$ & $0.001 \pm 0.017$ & $0.111^{+0.041}_{-0.044}$\\
OGLE-SMC-06 & \ra{00}{51}{56}{62} & \dec{-73}{22}{18}{1} & SMC720.26.2296                 & $18.205 \pm 0.010$ & $1.152 \pm 0.014$ & $0.045^{+0.065}_{-0.071}$\\
OGLE-SMC-07 & \ra{00}{50}{33}{66} & \dec{-73}{36}{27}{3} & SMC720.19.4610                 & $18.676 \pm 0.010$ & $0.771 \pm 0.014$ & $0.048^{+0.061}_{-0.070}$\\
\enddata
\tablecomments{The table provides the mean observed $I$-band brightness and $V-I$ color in the baseline. The reddening $E(V-I)$ toward each event was taken from \citet{skowron2021}.}
\end{deluxetable*}

Out of all analyzed light curves, only five met all selection cuts. These events constitute the statistical sample that we use for the calculation of the microlensing optical depth, event rate (Section~\ref{sec:tau}), and constraints on compact objects in dark matter (Section~\ref{sec:limits}). 
Events OGLE-SMC-03 and OGLE-SMC-04 were previously discovered by \citet{wyrzyk1} in the \mbox{OGLE-III} data, while OGLE-SMC-05, OGLE-SMC-06, and OGLE-SMC-07 are new discoveries. The equatorial coordinates of all events, along with their mean $I$-band magnitudes and $(V-I)$ colors in the baseline, are listed in Table~\ref{tab:events}. 

\movetabledown=2in
\begin{rotatetable}
\begin{deluxetable*}{crrrccrrr}
\tablecaption{Best-fit Parameters of Detected Events\label{tab:params}}
\tablehead{
\colhead{Event} & \colhead{$t_0$} & \colhead{$t_{\rm E}$ (d)} & \colhead{$u_0$} & \colhead{$\piEN$} & \colhead{$\piEE$} & \colhead{$I_{\rm s}$ (mag)} & \colhead{$f_{\rm s}$} & \colhead{$\chi^2/\mathrm{d.o.f.}$}}
\startdata
OGLE-SMC-03 & $2,454,476.09 \pm 0.10$ & $43.2^{+6.3}_{-5.3}$    & $0.150 \pm 0.026$         & \multicolumn{1}{c}{\dots} & \multicolumn{1}{c}{\dots} & $20.33 \pm 0.20$ & $1.33^{+0.27}_{-0.23}$ & 1364.0/1371 \\
OGLE-SMC-03 & $2,454,476.12 \pm 0.12$ & $41.8^{+5.7}_{-5.2}$    & $0.169 \pm 0.029$         & $+1.80 \pm 0.23$           & $+0.13 \pm 0.13$           & $20.18 \pm 0.21$ & $1.54^{+0.31}_{-0.28}$ & 1355.4/1371 \\
OGLE-SMC-03 & $2,454,476.19 \pm 0.11$ & $51.5^{+23.8}_{-13.8}$  & $0.117 \pm 0.052$         & $-0.91 \pm 0.56$          & $+0.28 \pm 0.23$           & $20.63^{+0.60}_{-0.44}$ & $1.01^{+0.45}_{-0.49}$ & 1360.8/1371 \\
OGLE-SMC-03 & $2,454,476.14 \pm 0.09$ & $36.8^{+5.3}_{-4.8}$    & $-0.166 \pm 0.029$        & $+2.01 \pm 0.27$           & $+0.10 \pm 0.15$           & $20.20 \pm 0.21$ & $1.51^{+0.30}_{-0.28}$ & 1355.1/1371 \\
OGLE-SMC-03 & $2,454,476.20 \pm 0.14$ & $59.3^{+40.9}_{-20.9}$  & $-0.111 \pm 0.056$        & $-1.04 \pm 0.60$          & $+0.34 \pm 0.27$           & $20.70^{+0.69}_{-0.49}$ & $0.95^{+0.48}_{-0.52}$ & 1360.8/1371 \\
OGLE-SMC-04 & $2,452,611.50 \pm 0.17$ & $18.6^{+2.0}_{-1.6}$    & $0.318 \pm 0.051$         & \multicolumn{1}{c}{\dots} & \multicolumn{1}{c}{\dots} & $18.66 \pm 0.24$ & $0.87^{+0.19}_{-0.17}$ & 1395.4/1383 \\
OGLE-SMC-05 & $2,456,112.40 \pm 1.75$ & $113.2^{+33.2}_{-25.3}$ & $0.159^{+0.081}_{-0.052}$ & \multicolumn{1}{c}{\dots} & \multicolumn{1}{c}{\dots} & $20.94 \pm 0.49$ & $0.75^{+0.43}_{-0.26}$ & 1226.1/1191 \\
OGLE-SMC-06 & $2,456,329.15 \pm 1.98$ & $214.7^{+21.0}_{-16.3}$ & $0.667 \pm 0.098$         & \multicolumn{1}{c}{\dots} & \multicolumn{1}{c}{\dots} & $18.45 \pm 0.27$ & $0.80^{+0.21}_{-0.18}$ & 1441.7/1405 \\
OGLE-SMC-06 & $2,456,366.85 \pm 5.98$ & $283.2^{+33.5}_{-32.3}$ & $0.349 \pm 0.076$         & $-0.0347 \pm 0.0057$      & $+0.0250 \pm 0.0100$       & $19.33 \pm 0.27$ & $0.35^{+0.10}_{-0.07}$ & 1411.6/1403 \\
OGLE-SMC-06 & $2,456,328.87 \pm 4.65$ & $238.3^{+18.7}_{-16.5}$ & $-0.728 \pm 0.076$        & $+0.0203 \pm 0.0199$       & $-0.0556 \pm 0.0133$      & $18.22 \pm 0.21$ & $0.99^{+0.14}_{-0.17}$ & 1423.7/1403 \\
OGLE-SMC-07 & $2,458,459.05 \pm 0.19$ & $20.0^{+3.0}_{-2.1}$    & $0.341 \pm 0.067$         & \multicolumn{1}{c}{\dots} & \multicolumn{1}{c}{\dots} & $18.84 \pm 0.30$ & $0.86 \pm 0.22$                   & 1403.5/1403 \\
OGLE-SMC-07 & $2,458,459.45 \pm 0.20$ & $18.7^{+2.8}_{-2.3}$    & $0.377 \pm 0.082$         & $+1.95 \pm 0.33$         & $+0.66 \pm 0.37$             & $18.68 \pm 0.33$ & $0.99 \pm 0.28$                   & 1380.2/1401 \\
OGLE-SMC-07 & $2,458,459.48 \pm 0.21$ & $19.3^{+2.9}_{-2.3}$    & $0.376 \pm 0.084$         & $+3.20 \pm 0.58$         & $-0.04 \pm 0.20$            & $18.69 \pm 0.34$ & $0.99 \pm 0.29$                   & 1380.0/1401 \\
OGLE-SMC-07 & $2,458,459.43 \pm 0.19$ & $17.8^{+2.7}_{-2.2}$    & $-0.376 \pm 0.082$        & $+2.15 \pm 0.37$         & $+0.56 \pm 0.36$             & $18.69 \pm 0.33$ & $0.99 \pm 0.29$                   & 1380.2/1401 \\
OGLE-SMC-07 & $2,458,459.43 \pm 0.18$ & $17.3^{+2.7}_{-2.2}$    & $-0.375 \pm 0.082$        & $+3.45 \pm 0.65$         & $-0.30 \pm 0.15$            & $18.69 \pm 0.33$ & $0.99 \pm 0.29$                   & 1380.0/1401 \\
\enddata
\end{deluxetable*}
\tablecomments{$I_{\rm s}$ is the observed source $I$-band magnitude.}
\end{rotatetable}

\begin{figure*}
\centering
\includegraphics[width=.49\textwidth]{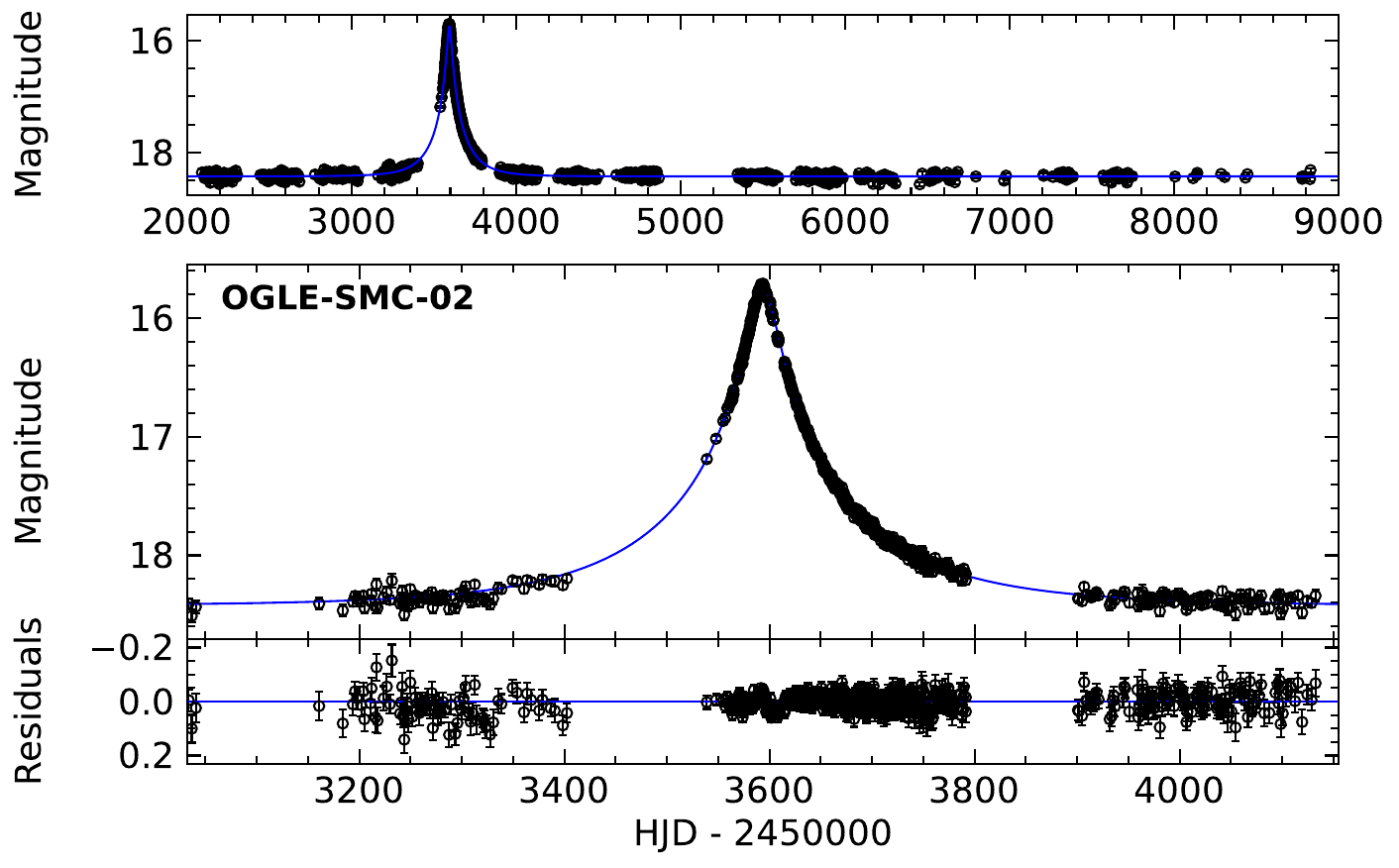}
\includegraphics[width=.49\textwidth]{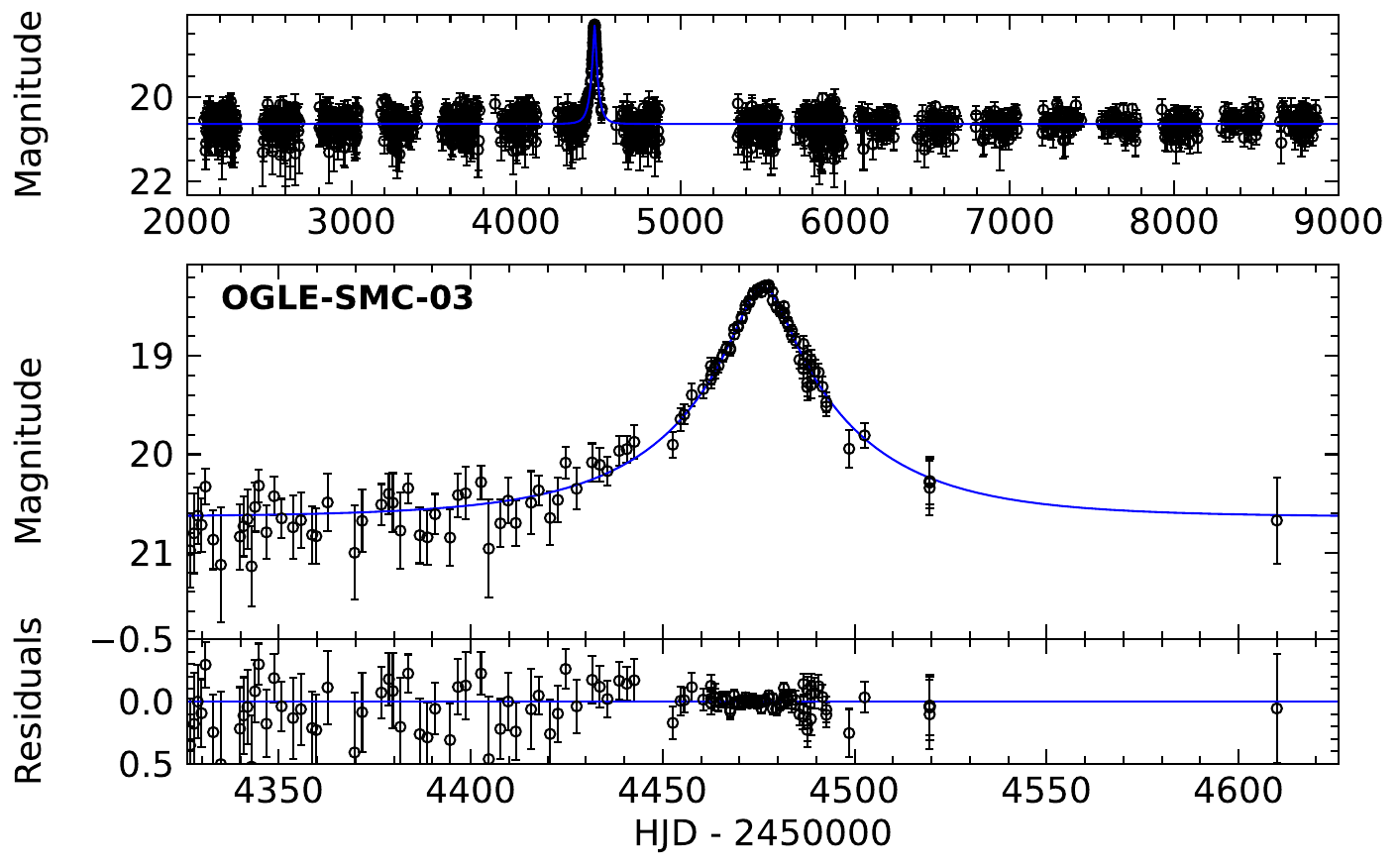}\\
\includegraphics[width=.49\textwidth]{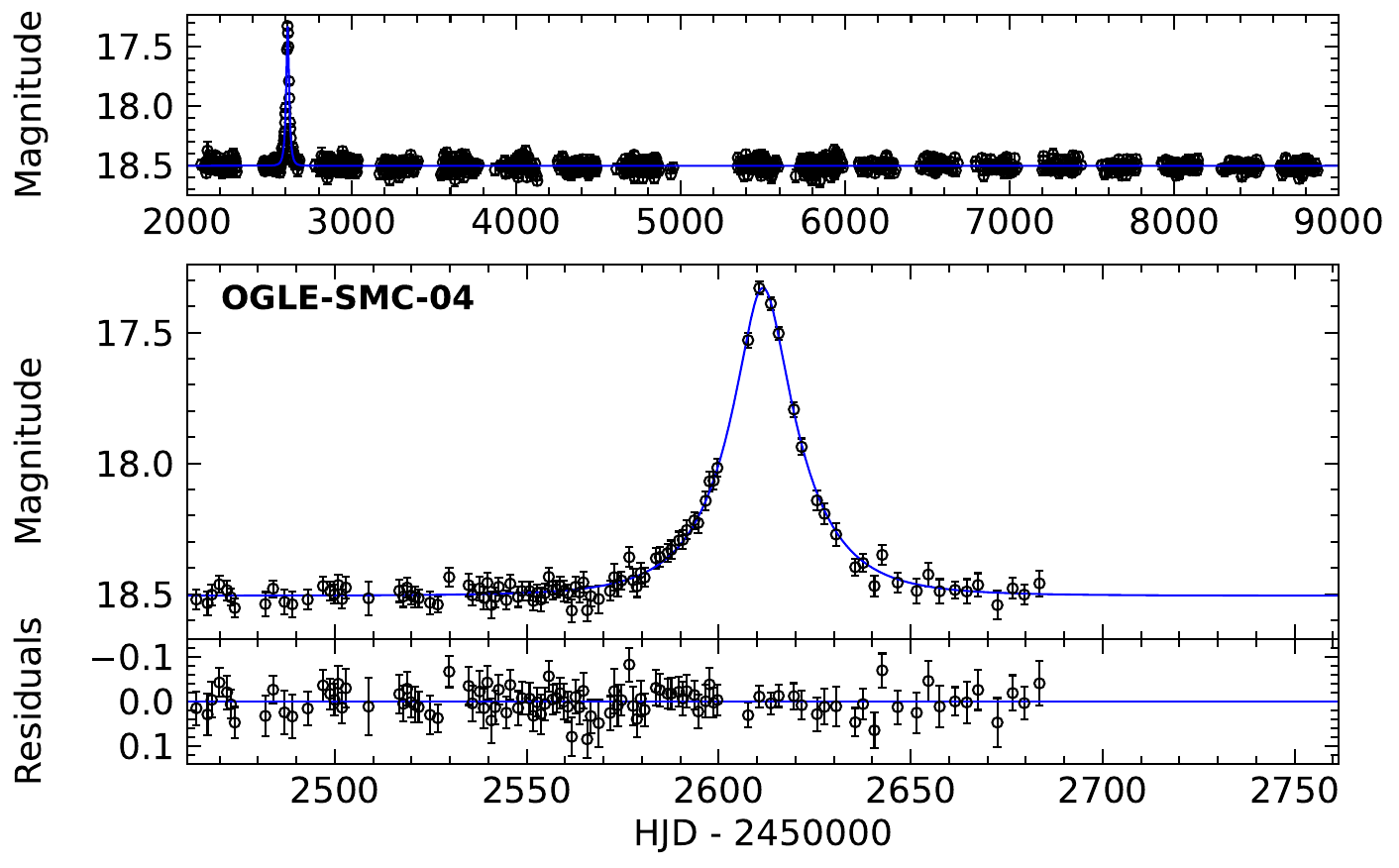}
\includegraphics[width=.49\textwidth]{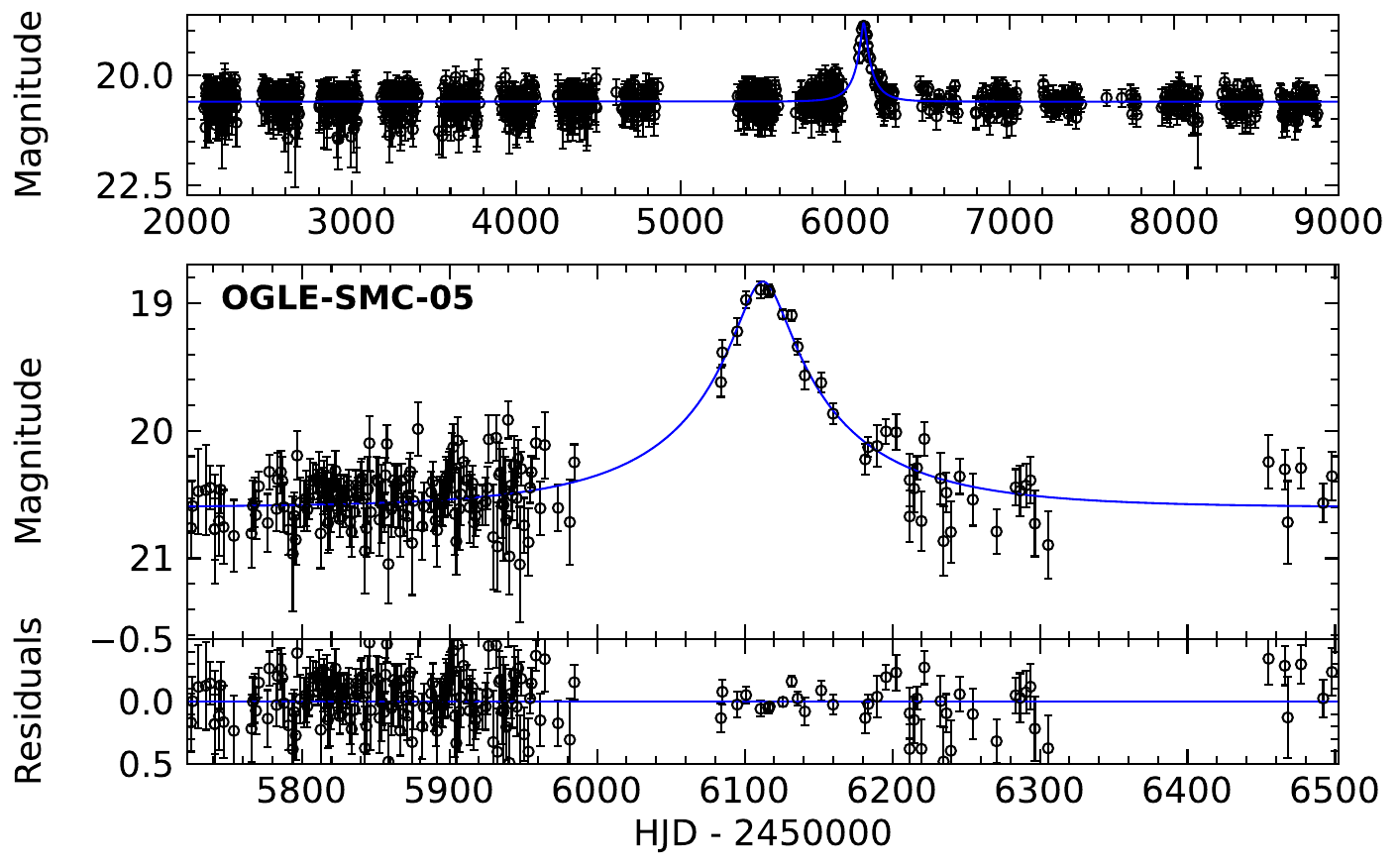}\\
\includegraphics[width=.49\textwidth]{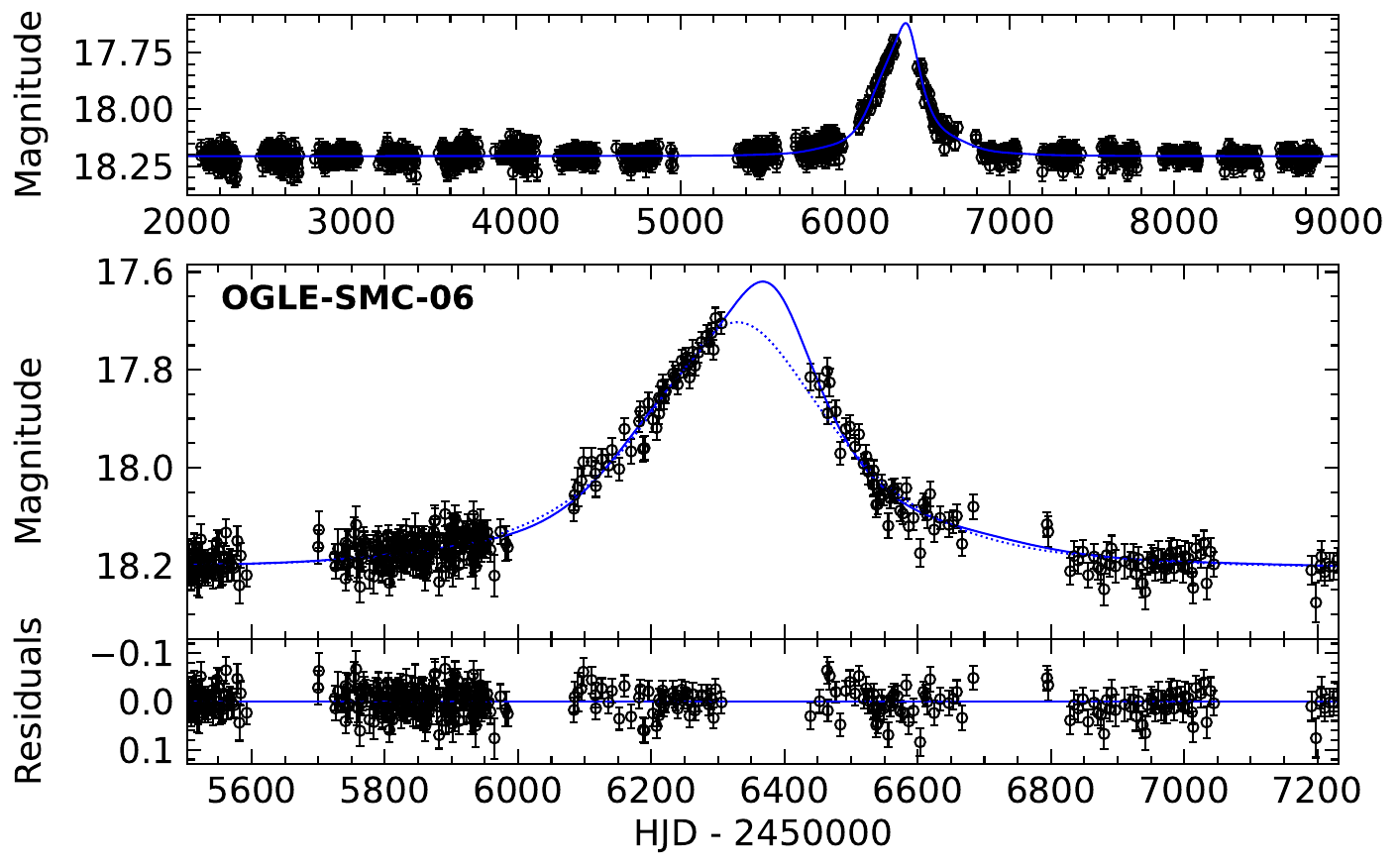}
\includegraphics[width=.49\textwidth]{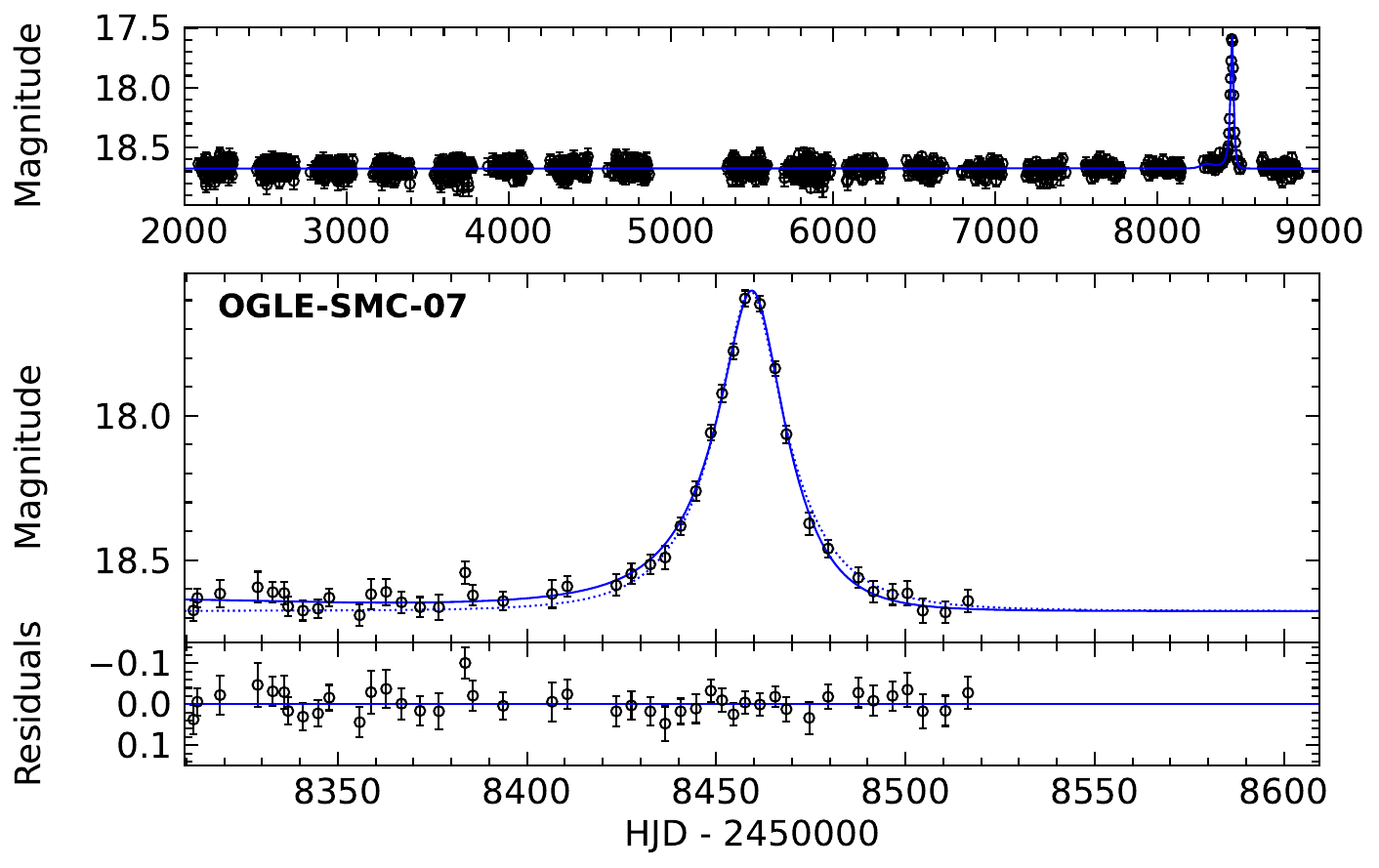}\\
\caption{$I$-band light curves of detected microlensing events. The solid blue line shows the best-fit microlensing model. The dotted lines for OGLE-SMC-06 and OGLE-SMC-07 present the best-fit PSPL models without microlensing parallax.}
\label{fig:light_curves}
\end{figure*}

\clearpage

For completeness, we visually inspected the light curves of all 25,973 objects selected in the first step of our criteria. We found only one additional event, OGLE-SMC-02, which was included in the sample of \citet{wyrzyk1}. This event is a known binary-lens event \citep{dong2007}, whose light curve significantly differs from that of a simple PSPL model. Hence, it did not satisfy all of our selection criteria. In addition, we checked the archival \mbox{OGLE-II} data (1997--2000), which were available for all events except OGLE-SMC-03 and OGLE-SMC-04. We also retrieved the archival MACHO light curves (1992--1999) for all events except OGLE-SMC-02 and OGLE-SMC-07. These additional data did not reveal any suspicious variability, which is consistent with the microlensing classification.

\section{Detected Events}
\label{sec:events}

Our final sample of microlensing events in the SMC consists of five objects. We modeled their light curves using the Markov Chain Monte Carlo (MCMC) techniques to refine the best-fit PSPL parameters and estimate their uncertainties. We used the same modeling techniques as described in \citetalias{mroz2024b}. The results are summarized in Table~\ref{tab:params}, where we provide the median values and the 68\% confidence intervals of the marginalized posterior distributions for all parameters. Note that OGLE-SMC-02 is excluded from this table because it is a binary-lens event that does not enter our statistical sample. The parameters of this event were provided by \citet{dong2007}. The light curves of analyzed events and the best-fit models are presented in Figure~\ref{fig:light_curves}.

In addition to PSPL models, we also included the annual microlensing parallax effect \citep{gould2004} in the fits. We found that the parallax is reliably measured for \mbox{OGLE-SMC-06}, for which we identified two possible degenerate solutions. These models, differing by the sign of $u_0$, are preferred over the standard PSPL model by $\Delta\chi^2=30.1$ and $\Delta\chi^2=18.0$ for the $u_0>0$ and $u_0<0$ solutions, respectively. This event, with an estimated timescale of $283.2^{+33.5}_{-32.3}$\,days ($u_0>0$) or $238.3^{+18.7}_{-16.5}$\,days ($u_0<0$), is the longest-timescale event identified toward the SMC to date. The long timescale of this event, along with a small value of the microlensing parallax, indicates that both the lens and the source are most likely located in the SMC.

A formally statistically significant signal of the microlensing parallax was also identified for \mbox{OGLE-SMC-07}. The fit improved by $\Delta\chi^2=23.5$ when including the parallax. This improvement mainly originates from the data points collected prior to the main event, in the time range $2,458,284 < \mathrm{HJD} < 2,458,411$, during which the event was magnified by $\approx 0.05$ mag compared to the mean baseline brightness. The parallax is also marginally ($\Delta\chi^2=8.6$) detected in the light curve of \mbox{OGLE-SMC-03}, as was originally noticed by \citet{wyrzyk1}. We identified four degenerate solutions for both events---two differing by the sign of $u_0$ \citep{smith2003} and two resulting from the jerk-parallax degeneracy \citep{gould2004}.
The parameters of both events (short timescales and large microlensing parallaxes) indicate that the lenses are located in the foreground Milky Way disk.

For three events, the $V$-band photometry collected during the magnified portion of the light curve was available. In these cases, we used model-independent regression to measure the color of the source star. We found the following values: $(V-I)_{\rm s}=0.920 \pm 0.003$\,mag for OGLE-SMC-02, $(V-I)_{\rm s} = 0.078 \pm 0.017$\,mag for OGLE-SMC-03, and $(V-I)_{\rm s}=1.214 \pm 0.009$\,mag for OGLE-SMC-06. These values are quite similar to the baseline colors (see Table~\ref{tab:events}), which confirms that the lens does not substantially contribute to the event's brightness or that the color of the lens is similar to that of the source star. 

\begin{figure}
\centering
\includegraphics[width=.5\textwidth]{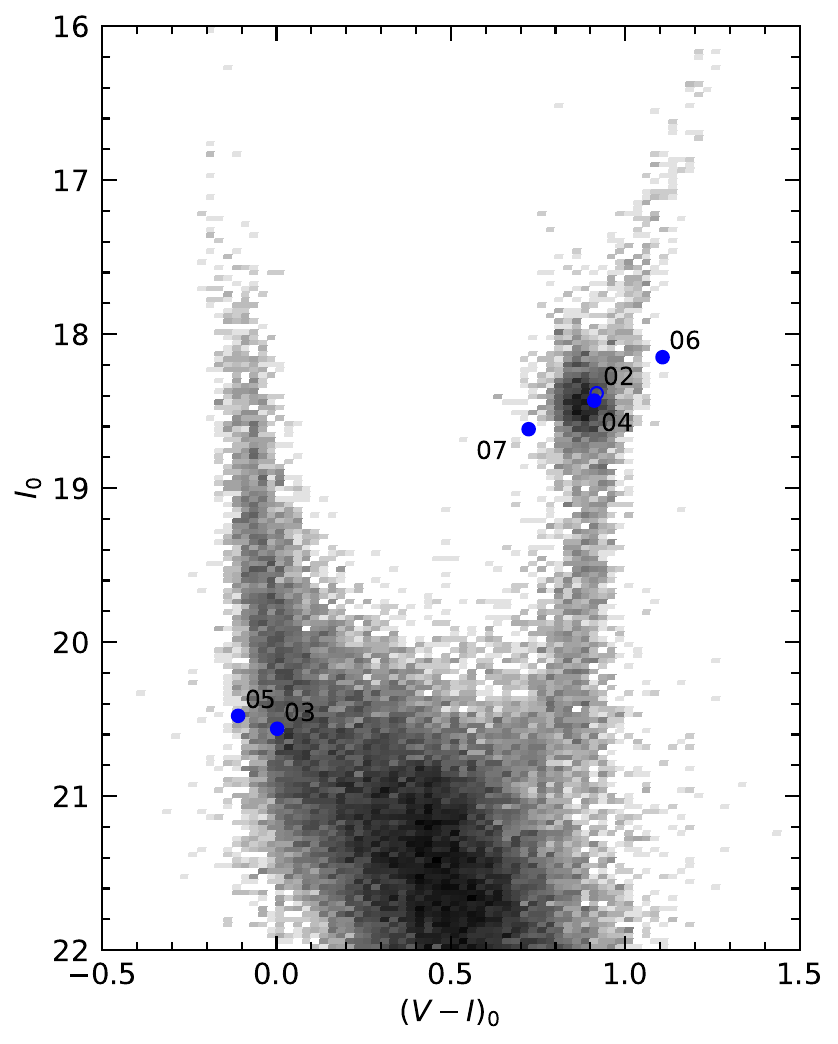}
\caption{De-reddened color--magnitude diagram of the field SMC726.28. Blue circles mark the positions of the detected microlensing events in the baseline.}
\label{fig:cmd}
\end{figure}

The positions of all detected events in the de-reddened CMD are presented as blue dots in Figure~\ref{fig:cmd}. The background CMD shows the locations of stars in the field SMC726.28, and blue dots mark the positions of the discovered events in the baseline:
\begin{align}
\begin{split}
I_0 &= I_{\rm base} - 1.217 E(V-I), \\
(V-I)_0 &= (V-I)_{\rm base} - E(V-I),
\end{split}
\end{align}
where we adopted the $I$-band extinction to reddening ratio of $A_I / E(V-I)=1.217$ from \citet{schlafly2011}.
The reddening toward each event was estimated using optical reddening maps by \citet{skowron2021}. Four of the detected events have sources situated close to the red clump in the CMD, while the positions of the other two events indicate main-sequence sources.

Among the newly identified events, the classification of OGLE-SMC-05 as a microlensing event appears to be the least certain. Although the data can be reasonably well fitted with the PSPL model, a rising portion of the light curve is missing. This raises the possibility that the object might be a supernova. Additionally, as no $V$-band observations were secured during the magnified portion of the light curve, we cannot confirm whether it is achromatic, as would be expected for microlensing events. On a positive note, the source looks stellar in the reference image, and there are no galaxies located near this event.

\subsection{Comparison with Previous Searches for Microlensing Events toward the SMC}

The analyzed data set, by construction, contains the majority of \mbox{OGLE-III} light curves that were previously studied by \citet{wyrzyk1}, who reported the discovery of three microlensing events. We were able to successfully recover two of these events (OGLE-SMC-03 and OGLE-SMC-04). Their best-fit parameters are very similar to those measured by \citet{wyrzyk1} (Table~\ref{tab:params}). However, one event (OGLE-SMC-02) was not selected by our automated pipeline because it did not pass all selection cuts, in particular those related to the PSPL fit quality. That is not surprising because this event is known to be a weak (non-caustic-crossing) binary-lens event whose light curve significantly deviates from a simple PSPL model \citep{dong2007}. For completeness, we include the light curve of OGLE-SMC-02 in the upper-left panel of Figure~\ref{fig:light_curves}.

We also examined the light curves of other SMC microlensing candidates that have been reported in the literature. Thanks to the extended time span of observations, we were able to search for additional outbursts or variability that could disprove the microlensing classification. For example, \citet{wyrzyk3} reported the discovery of one microlensing candidate (called OGLE-SMC-01) in the OGLE-II data. This event was considered to be ``rather weak'' because of a sparsely sampled light curve and an unusually blue color in the baseline, which placed the event in a relatively empty region of the CMD between the main sequence and the red giant branch. Although \citet{wyrzyk3} noted that the star did not show any additional outbursts in the \mbox{OGLE-III} data (2001--2009), the \mbox{OGLE-IV} data reveal a second outburst occurring between January and November 2016, demonstrating that this object is not a microlensing event.

\citet{tisserand2007} reported the detection of two microlensing candidates in the SMC. The first event, EROS2-SMC-1 (a.k.a. MACHO-97-SMC-1), was discovered by \citet{alcock1997_smc} and \citet{palanque1998}. The full OGLE light curve of this event does not reveal any additional outbursts. However, the source star exhibits sinusoidal variability with a period of $5.12505(3)$ days. This derived period is close to those measured by \citet{alcock1997_smc}, \citet{udalski1997b}, and \citet{palanque1998}. The source is likely a binary system with an orbital period twice as long as the photometric period. Another candidate reported by \citet{tisserand2007}, EROS2-SMC-5, underwent a second outburst in November 2017, which is inconsistent with a microlensing interpretation. Additionally, the caustic-crossing event MACHO-98-SMC-1 \citep{alcock1999,afonso2000} is located in a gap between the CCD detectors of the \mbox{OGLE-IV} camera, and therefore, no recent data are available for this event.

\section{Number of Source Stars}
\label{sec:stars}

The number of stars that are detected on OGLE reference images differs from the number of source stars that might give rise to microlensing events. This difference arises for two main reasons. First, the resolution of ground-based images is limited by seeing, which can cause two or more nearby stars to appear blended and unresolved. Second, the probability that nearby foreground stars are microlensed is virtually zero; such stars therefore do not significantly contribute to the observed number of microlensing events.

We used the methods developed in \citetalias{mroz2024b} to estimate the number of possible source stars. In short, we first compared the surface density of stars brighter than a certain limiting magnitude on the OGLE reference images ($\Sigma^{\rm OGLE}$) with the corresponding surface density $\Sigma^{\rm HST}$ derived from deep Hubble Space Telescope (HST) observations of the same fields. We used the HST images compiled and reduced by \citet{holtzman2006}. These HST images served as a ``ground truth.'' The data set created by \citet{holtzman2006} includes nine ``priority 1'' fields in the SMC---those with HST observations in the F555W and F814W filters and exposure times longer than 500\,s. Two of these fields contain stellar clusters; for the remaining seven, we calculated the correction factor (CF), defined as $\mathrm{CF}=\Sigma^{\rm HST}/\Sigma^{\rm OGLE}$, for stars brighter than $I=21$, $I=21.5$, and $I=22$\,mag. 

We found that these correction factors were very similar to those derived in \citetalias{mroz2024b} for the LMC. This conclusion is further supported by analyses of simulations carried out by \citet{wyrzykowski2011} and \citet{wyrzyk1}, which show that the correction factors for the LMC and SMC are consistent within 5\%. Because the number of SMC fields included in the \citet{holtzman2006} database is much smaller than the number of LMC fields (64), we adopted the correction factors derived in \citetalias{mroz2024b}.

We then estimated the surface density of foreground stars by fitting the following formula:
\begin{equation}
\log \Sigma^{\rm foreground} = \xi + \eta (|b| - 30^{\circ})
\end{equation}
to stellar surface densities calculated in the outer SMC fields (those with $b<-52^{\circ}$ or $b>-37.5^{\circ}$). Here, $b$ is the Galactic latitude, and $\xi$ and $\eta$ are numerical coefficients. We obtained $\xi=0.783 \pm 0.049$, $\eta=-0.0071 \pm 0.0024$ for stars brighter than $I=21$ mag; $\xi=0.805 \pm 0.074$, $\eta=-0.0051 \pm 0.0036$ for stars brighter than $I=21.5$ mag; $\xi=0.823 \pm 0.081$, $\eta=-0.0047 \pm 0.0040$ for stars brighter than $I=22$ mag.

\begin{figure}
\centering
\includegraphics[width=.5\textwidth]{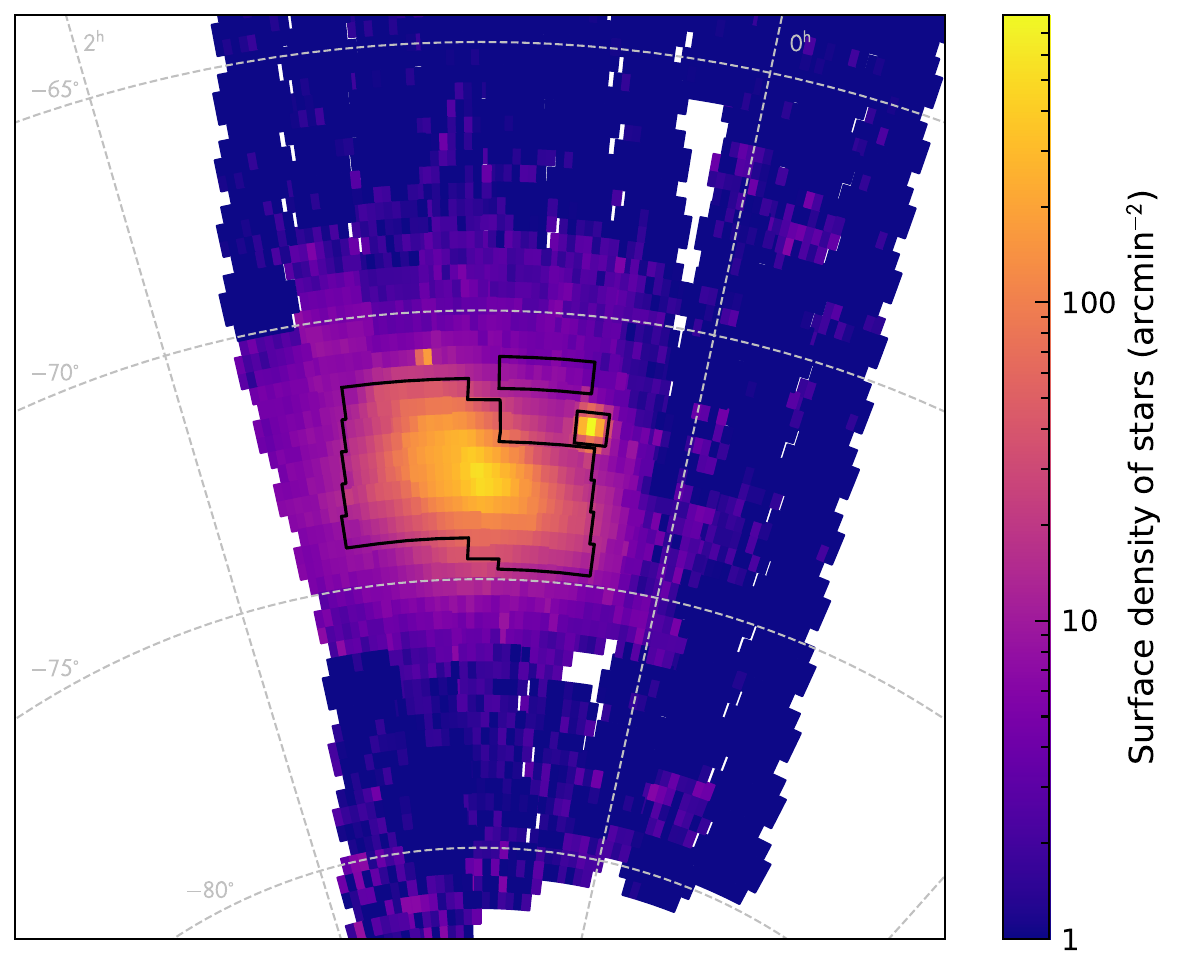}
\caption{Surface density of source stars brighter than $I=21$\,mag in the analyzed fields. The black polygons mark the area observed by \mbox{OGLE-III}. The map is presented in an equal-area Hammer projection centered on $\mathrm{R.A.}=0^{\rm h}51^{\rm m}$, $\mathrm{Decl.}=-73.1^{\circ}$.}
\label{fig:stars}
\end{figure}

To calculate the number of source stars, we multiplied the number of stars detected on the OGLE reference images by the correction factor and subtracted the estimated number of foreground stars. The results are summarized in Table~\ref{tab:stars}, which presents the total number of stars ($N_{\rm tot}$) and the estimated number of SMC source stars ($N_{\rm corr}$) that are brighter than $I=21$, $I=21.5$, and $I=22$\,mag in each subfield. Overall, we estimate that \mbox{OGLE-IV} monitored about $6.32 \pm 0.19$ million source stars brighter than $I=21$\,mag, $9.69 \pm 0.68$ million source stars brighter than $I=21.5$\,mag, and $12.4 \pm 1.4$ million source stars brighter than $I=22$\,mag. The error bars represent the accuracy of determination of star counts, which we adopted from \citetalias{mroz2024b}. The surface density of source stars brighter than $I=21$\,mag is presented in Figure~\ref{fig:stars}.

We compared our results with the number of sources reported by \citet{wyrzyk1}, who found about 5.97~million source stars brighter than $I=21$\,mag in the \mbox{OGLE-III} SMC fields. We estimate that about 91.6\% of them (that is, about 5.47 million) should also be detected in the \mbox{OGLE-IV} reference images, the remaining 8.4\% are located in the gaps between the CCD detectors or between the \mbox{OGLE-IV} fields. If we restrict our sample to stars brighter than $I=21$\,mag and to those located in the \mbox{OGLE-III} fields, we found a total of 5.47 million stars (5.16 million after removing foreground contamination), in excellent agreement with the results presented by \citet{wyrzyk1}.

\begin{deluxetable*}{rrrrrrrrr}
\tablecaption{Number of Microlensing Source Stars Observed in \mbox{OGLE-IV} Fields\label{tab:stars}}
\tablehead{
\colhead{Field} & \colhead{R.A. (deg)} & \colhead{Decl. (deg)} & \colhead{$N_{\rm total}^{\rm 21\,mag}$} & \colhead{$N_{\rm corr}^{\rm 21\,mag}$} & \colhead{$N_{\rm total}^{\rm 21.5\,mag}$} & \colhead{$N_{\rm corr}^{\rm 21.5\,mag}$} & \colhead{$N_{\rm total}^{\rm 22\,mag}$} & \colhead{$N_{\rm corr}^{\rm 22\,mag}$}}
\startdata
SMC700.01 & 5.1554 & --69.9144 & 1188 & 413 & 1697 &  817 & 1895 &  963 \\
SMC700.02 & 4.7161 & --69.9144 & 1168 & 393 & 1606 &  725 & 1750 &  817 \\
SMC700.03 & 4.2768 & --69.9144 & 1229 & 453 & 1640 &  758 & 1812 &  878 \\
SMC700.04 & 3.8376 & --69.9144 & 1231 & 455 & 1732 &  850 & 1895 &  961 \\
SMC700.05 & 3.3983 & --69.9144 & 1142 & 365 & 1571 &  688 & 1699 &  764 \\
SMC700.06 & 2.9591 & --69.9144 & 1233 & 456 & 1637 &  754 & 1770 &  834 \\
SMC700.07 & 2.5198 & --69.9144 &  952 & 174 & 1271 &  387 & 1390 &  454 \\
SMC700.08 & 5.5946 & --69.5930 & 1419 & 649 & 1881 & 1003 & 2033 & 1103 \\
SMC700.09 & 5.1554 & --69.5930 & 1278 & 507 & 1824 &  946 & 2066 & 1136 \\
SMC700.10 & 4.7161 & --69.5930 & 1150 & 379 & 1593 &  714 & 1700 &  770 \\
\multicolumn{1}{c}{\dots} & \multicolumn{1}{c}{\dots} & \multicolumn{1}{c}{\dots} & \multicolumn{1}{c}{\dots} & \multicolumn{1}{c}{\dots} & \multicolumn{1}{c}{\dots} & \multicolumn{1}{c}{\dots} & \multicolumn{1}{c}{\dots} & \multicolumn{1}{c}{\dots}\\
Total (million) & \multicolumn{1}{c}{\dots} & \multicolumn{1}{c}{\dots} & 9.606 & 6.316 & 13.364 & 9.689 & 16.230 & 12.351\\
\enddata
\tablecomments{(This table is available in its entirety in machine-readable form.)}
\end{deluxetable*}

\begin{figure}
\centering
\includegraphics[width=.5\textwidth]{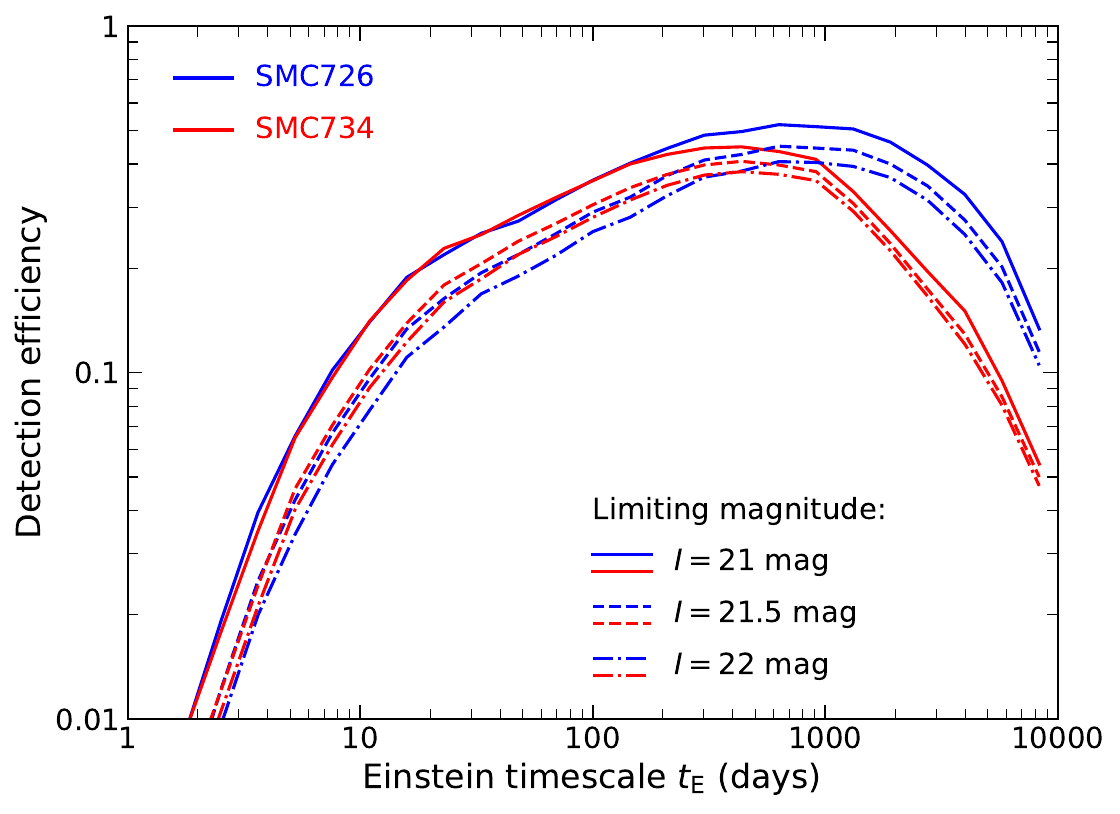}
\caption{Example detection efficiency curves for representative fields SMC726 and SMC734. The solid, dashed, and dash--dotted lines represent detection efficiencies averaged for sources brighter than $I=21$, $I=21.5$, and $I=22$~mag, respectively.}
\label{fig:eff}
\end{figure}

\section{Detection Efficiency}
\label{sec:eff}

We used methods developed in \citetalias{mroz2024b} to calculate the event detection efficiency as a function of the event timescale. A detailed description of our calculations can be found in \citetalias{mroz2024b}. In brief, we created synthetic light curves of microlensing events by injecting artificial signals into the light curves of randomly selected stars from the database. This approach allows us to preserve the sampling of the original light curves, as well as the variability and noise present in the actual data, which would be challenging to simulate otherwise. We applied exactly the same selection criteria outlined in Section~\ref{sec:search} to the simulated light curves and ultimately measured the fraction of synthetic events that satisfied all selection criteria. The resulting detection efficiency was subsequently lowered by 10\% to account for possible binary-lens events \citep{wyrzyk1, mroz2024b} that may be missed by our automated detection pipeline.

The parameters of events were drawn from physically motivated distributions. The time of the maximum was sampled from a uniform distribution within the time ranges $2,452,000 \leq t_0 \leq 2,459,000$ (for objects observed during both \mbox{OGLE-III} and \mbox{OGLE-IV}) and $2,455,000 \leq t_0 \leq 2,459,000$ (for stars observed during \mbox{OGLE-IV} only). The impact parameters were drawn from a uniform distribution from the range $0 \leq u_0 \leq 1$, while the Einstein timescales were drawn from a log-uniform distribution from the range $1 \leq t_{\rm E} \leq 10^4$\,days. For the blending parameter $f_{\rm s}$, we used an empirical distribution created by matching stars detected on OGLE reference images with those detected by HST. 

We simulated 10,000 light curves per detector (totaling 320,000 light curves per field) and calculated the detection efficiency across 25 identical bins in $\log\tE$. For each field, we calculated detection efficiencies averaged over sources brighter than $I=21$, $I=21.5$, and $I=22$~mag. Figure~\ref{fig:eff} shows example detection efficiency curves for representative fields SMC726 (observed during both \mbox{OGLE-III} and \mbox{OGLE-IV}) and SMC734 (observed during \mbox{OGLE-IV} only).

\section{Optical Depth and Event Rate}
\label{sec:tau}

Given the statistical sample of five microlensing events toward the SMC (Section~\ref{sec:search}), event detection efficiency $\varepsilon(\tE)$ calculated in Section~\ref{sec:eff}, and the total number of observed source stars $N_{\rm s}$(Section~\ref{sec:stars}), we can compute the microlensing optical depth $\tau$ and event rate $\Gamma$ using the following equations:
\begin{align}
\begin{split}
\tau &= \frac{\pi}{2 N_{\rm s}}\sum_i\frac{1}{\Delta T_i}\frac{t_{{\rm E},i}}{\varepsilon(t_{{\rm E},i})},\\
\Gamma &= \frac{1}{N_{\rm s}}\sum_i\frac{1}{\Delta T_i}\frac{1}{\varepsilon(t_{{\rm E},i})},
\end{split}
\end{align}
where $\Delta T$ is the duration of observations. For fields observed during both \mbox{OGLE-III} and \mbox{OGLE-IV} surveys, we use $\Delta T=7000$~days, while for field observed only during \mbox{OGLE-IV}, we use $\Delta T=4000$~days. To account for uncertainties on the event timescale, we replace the terms $t_{{\rm E},i}/\varepsilon(t_{{\rm E},i})$ and $1/\varepsilon(t_{{\rm E},i})$ with their means calculated from the posterior distribution of $\tE$ (see \citetalias{mroz2024b} for more details).

For our calculations, we use the five events that met all selection criteria outlined in Section~\ref{sec:search}, alongside the number of sources brighter than $I=22$~mag ($N_{\rm s} = (12.4 \pm 1.4) \times 10^6$) and the detection efficiencies calculated for sources brighter than this limit. The total microlensing optical depth toward the \mbox{OGLE-IV} SMC fields is found to be $\tau = (0.32 \pm 0.18) \times 10^{-7}$, while the event rate is determined to be $\Gamma = (1.18 \pm 0.57) \times 10^{-7}\,\mathrm{yr}^{-1}\,\mathrm{star}^{-1}$. The uncertainties of these quantities were calculated using methods proposed by \citet{han1995_stat} and \citet{mroz2019b}. These error bars reflect only the uncertainties caused by the small number of events and by the uncertainties in Einstein timescales. There is an additional source of error (approximately 11\%) related to the accuracy of measuring the number of source stars.  The contribution of each event to the total optical depth and event rate is detailed in Table~\ref{tab:tau}.

\begin{deluxetable*}{lccccc}
\tablecaption{Microlensing Optical Depth and Event Rate Toward the LMC\label{tab:tau}}
\tablehead{
\colhead{Event} & \colhead{$\langle\varepsilon(\tE)\rangle$} & \colhead{$\langle1/\varepsilon(\tE)\rangle$} & \colhead{$\langle\tE/\varepsilon(\tE)\rangle$} & \colhead{$\tau$} & \colhead{$\Gamma$}\\
\colhead{} & \colhead{} & \colhead{} & \colhead{(days)} & \colhead{($10^{-7}$)} & \colhead{$(10^{-7}\,\mathrm{yr}^{-1})$}}
\startdata
OGLE-SMC-03 & 0.1848 & 5.430 & 235.76 & 0.0428 & 0.2294 \\
OGLE-SMC-04 & 0.1205 & 8.379 & 156.17 & 0.0284 & 0.3540 \\
OGLE-SMC-05 & 0.2635 & 3.827 & 439.51 & 0.0799 & 0.1617 \\ 
OGLE-SMC-06 & 0.3599 & 2.783 & 787.51 & 0.1431 & 0.1176 \\
OGLE-SMC-07 & 0.1336 & 7.593 & 153.36 & 0.0279 & 0.3208 \\
\hline
Total: & \dots & \dots & \dots & $0.32 \pm 0.18$ & $1.18 \pm 0.57$\\
\enddata
\end{deluxetable*}

\section{Expected Number of Events from Stellar Populations}
\label{sec:sl}

In this section, we present simple models of microlensing by stellar populations (stars, brown dwarfs, and stellar remnants) in the SMC and the foreground Milky Way disk, and demonstrate that they can explain the number and timescales of observed microlensing events toward the SMC. The differential event rate toward a source star at a distance of $\Ds$ is given by the formula
\begin{equation}
\frac{d^4\Gamma}{d\Dl dM d^2\mu}=2\thetaE\mu \Dl^2 \nu(\Dl) f(\muvec)g(M),
\end{equation}
which was first introduced by \citet{batista2011}. Here, $M$ is the mass of the lens, $f(\muvec)$ is the 2D probability function for a given source--lens relative proper motion $\muvec$, $\thetaE$ is the angular Einstein radius, $\nu(\Dl)$ is the number density of lenses at a given distance $\Dl$, and $g(M)$ is the mass function of lenses. The total event rate in a given direction can be expressed as the following integral:
\begin{equation}
\Gamma = \frac{\int_0^{\infty}\rho(\Ds)\Ds^{2-2\beta}d\Ds \int_0^{\Ds} \frac{d^4\Gamma}{d\Dl dM d^2\mu} d\Dl dM d^2\mu}{\int_0^{\infty}\rho(\Ds)\Ds^{2-2\beta}d\Ds},
\label{eq:gamma}
\end{equation}
where $\rho(\Ds)$ is the density of sources, and the coefficient $\beta$ quantifies how the number of observed sources changes with their distance \citep{kiraga1994}. It is generally assumed that $0 \leq \beta \leq 1$. Here, we adopt $\beta=0$ as a fiducial value, but we demonstrate below that the results weakly depend on the choice of this parameter.

The integrals in Equation~(\ref{eq:gamma}) are evaluated numerically using the Monte Carlo integration technique. For example, to calculate the multidimensional integral in the numerator of Equation~(\ref{eq:gamma}), we draw a random sample of $10^8$ points from a 5D cube $D_{\rm l, min} \leq D_{\rm l} \leq D_{\rm l, max}$, $D_{\rm s, min} \leq D_{\rm s} \leq D_{\rm s, max}$, $M_{\rm min} \leq M \leq M_{\rm max}$, $\mu_{\rm min} \leq \mu_{x,y} \leq \mu_{\rm max}$, evaluate the integrand at each point, and then calculate its mean value. To speed up the calculations, we change from the variable $M$ to $\log M$. All integrals are evaluated with precision better than 1\%.

As a byproduct of the Monte Carlo integration, we also evaluate the expected event timescale distribution $d\Gamma/d\log\tE$ (where $\tE = \thetaE/\mu$). The total number of expected events in a given field can be found by
\begin{equation}
N_{\rm exp} = N_{\rm s} \Delta T \int \frac{d\Gamma}{d\log\tE}\varepsilon(\tE) d\log\tE,
\label{eq:n_exp}
\end{equation}
where $N_s$ is the number of source stars, $\Delta T$ is the duration of the experiment ($\Delta T=7000$ days or $\Delta T=4000$ days, depending on whether the field was observed during \mbox{OGLE-III}, see Section~\ref{sec:tau}), and $\varepsilon(\tE)$ is the event detection efficiency measured in that field.

\subsection{SMC Self-Lensing Models}

We analyze two SMC self-lensing models. The first one is based on the model proposed by \citet{calchi_novati2013}, in which the total stellar mass (within 5\,kpc of the SMC center) is taken to be $M_* = 1.0 \times 10^9\,M_{\odot}$ \citep{bekki2009}. However, this quantity is by no means well determined, and \citet{calchi_novati2013} argued that $M_*$ is known with an accuracy of up to a factor of 2. For example, \citet{mcconnachie2012} reported $M_* = 4.6 \times 10^8\,M_{\odot}$. Because the derived microlensing event rate is directly proportional to the total stellar mass, the former quantity is also accurate up to a factor of 2.

The model of stellar populations comprises two components, the old and young stellar populations, whose mass ratio is assumed to be 3:2 \citep{bekki2009}. The young stellar population is approximated with a 3D Gaussian profile with length scales of 0.8, 3.5, and 1.3\,kpc. On the other hand, the young population is described using the Gaussian profile along the line of sight (length scale of 2.1\,kpc) and a smoother exponential profile in the plane of the sky (with length scales of 0.8 and 1.2\,kpc). Following \citet{calchi_novati2013}, we adopt the coordinates of the SMC center as reported by \citet{gonidakis2009} ($\alpha=0^{\rm h}51^{\rm m}$, $\delta=-73.1^{\circ}$, J2000), and the SMC distance of 61.5\,kpc. \citet{graczyk2020} recently reported a slightly greater geometric distance to the SMC center of $62.44 \pm 0.47$\,kpc, but we found that the choice of this quantity has a negligible effect on our results. The inclination and position angles of the reference frames of the young and old stellar populations are taken from \citet{calchi_novati2013}.

An alternative self-lensing model is based on the methodology proposed by \citet{mroz_poleski2018}. In this approach, the spatial distribution of both young and old stellar populations is approximated using a Gaussian mixture model, in which the total density is represented as a sum of 3D Gaussians with different weights. That enables us to more accurately model the irregular shape of this galaxy. We used the \textsc{scikit-learn} package \citep{scikit-learn} to fit 32 Gaussian components to the 3D distributions of classical Cepheids \citep{jacyszyn2016} and RR Lyrae stars \citep{jacyszyn2017} in the SMC. These stars are intended to represent the young and old stellar populations, respectively. The SMC center, as adopted by \citet{jacyszyn2016} and \citet{jacyszyn2017}, has equatorial coordinates of $\alpha=1^{\rm h}05^{\rm m}$, $\delta=-72^{\circ}25.2'$ (J2000). We assume a total stellar mass of $M_* = 1.0 \times 10^9\,M_{\odot}$, with a mass ratio of old to young stellar population set as 3:2.

\begin{figure*}
\includegraphics[width=\textwidth]{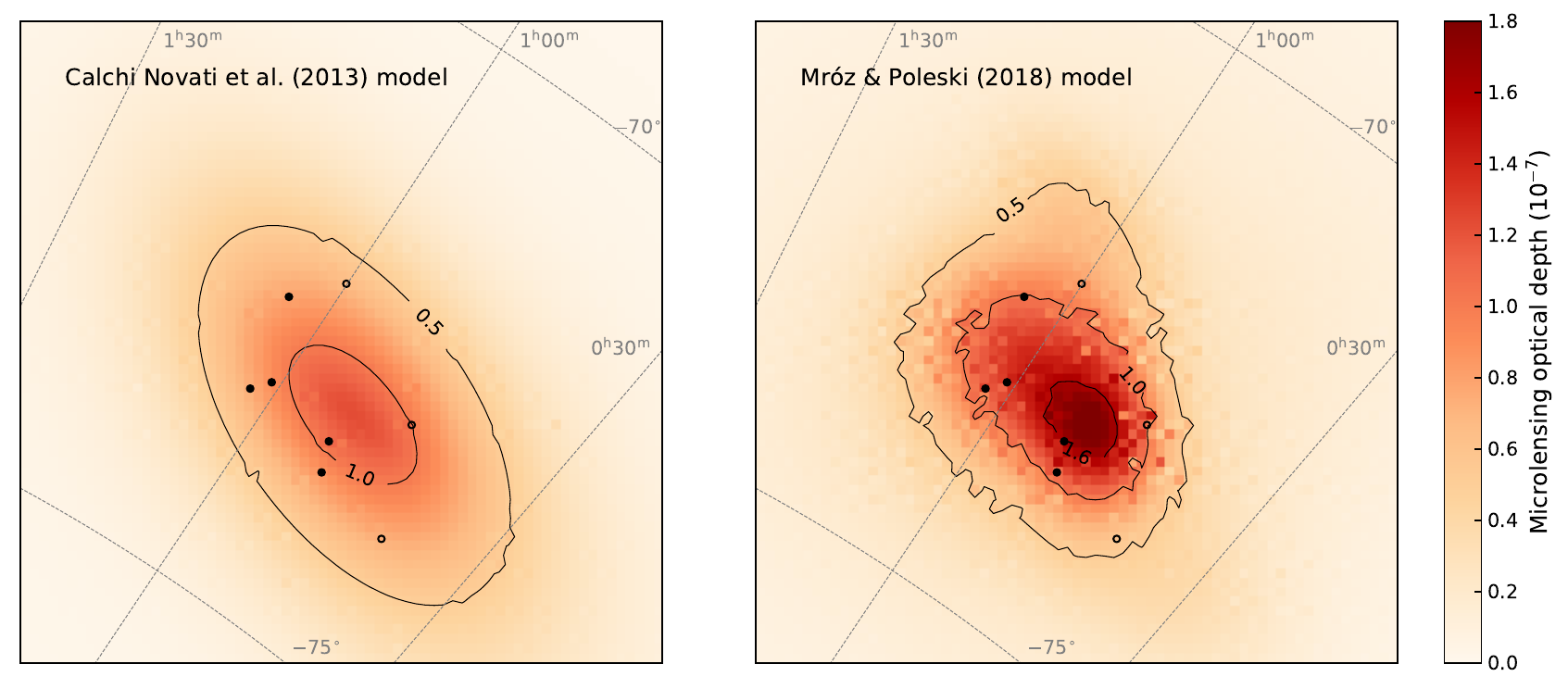}
\caption{Contours of the microlensing optical depth for two self-lensing SMC models: \citet{calchi_novati2013} (left panel) and \citet{mroz_poleski2018} (right). The filled circles indicate the positions of five microlensing events that are included in the statistical sample analyzed in this paper, while the empty circles represent other known microlensing events.}
\label{fig:tau}
\end{figure*}

The contours of the microlensing optical depth in both models are presented in Figure~\ref{fig:tau}. The \citet{calchi_novati2013} model has a peak value of $1.3 \times 10^{-7}$, whereas the Gaussian mixture model \citep{mroz_poleski2018} predicts a higher maximal value of $1.9 \times 10^{-7}$. The optical depth contours in the latter model are not symmetric, and reflect the irregular distribution of classical Cepheids in the SMC.

For both stellar distribution models, we use the empirical model of the SMC kinematics that was derived using the Gaia DR3 data (see the Appendix~\ref{sec:smc_pm}). We assume that the mass function of lenses is identical to that measured by \citet{kroupa2001}. For the evaluation of integrals in Equation~(\ref{eq:gamma}), we adopt $D_{\rm l, min}=D_{\rm s, min}=48$\,kpc, $D_{\rm l, max}=D_{\rm s, max}=76$\,kpc, $\mu_{\rm min}=-0.5$\,mas\,yr$^{-1}$, $\mu_{\rm max}=0.5$\,mas\,yr$^{-1}$, $M_{\rm min}=0.01\,M_{\odot}$, and $M_{\rm max}=10\,M_{\odot}$. The event rate was evaluated at the center of each field (Table~2). For four central fields (SMC719, SMC720, SMC725, and SMC726), the integrals in Equation~(\ref{eq:gamma}) were evaluated at the centers of all 32 subfields (that is, separately for each CCD detector), and then averaged using the expected number of sources (denominator in Equation~(\ref{eq:gamma})) as a weight.

\subsection{Milky Way Disk}

We consider two possible Milky Way disk models \citep{han_gould2003,cautun2020}, identical to those used in \citetalias{mroz2024a} for calculating the expected number of microlensing events toward the LMC. However, because the SMC center ($b=-44^{\circ}$) is located at higher Galactic latitude than the LMC center ($b=-33^{\circ}$), the expected event rate from foreground Milky Way lenses is smaller by $\approx 30$\%. Both Milky Way disk models are composed of two stellar populations (thin and thick disk) that are described by the double exponential profiles:
\begin{equation}
\rho(R,z) = \rho_0 \exp\left(-\frac{R}{h_R}\right) \exp\left(-\frac{|z|}{h_z}\right),
\end{equation}
where $R$ is the Galactocentric distance, $z$ is the distance from the Milky Way plane, and $h_R$ and $h_z$ are the radial scale length and scale height, respectively. The model parameters are shown in Table~\ref{tab:mw_disk}. We assume that stars in the disk are moving with the velocity given by the Milky Way rotation curve (Section~\ref{sec:limits}) with the dispersion of 30\,km\,s$^{-1}$, and their mass function follows that of \citet{kroupa2001}.

For the numerical calculation of the event rate (Equation~(\ref{eq:gamma})) we use the following limits $D_{\rm s, min}=48$\,kpc, $D_{\rm s, max}=76$\,kpc, $D_{\rm l, min}=0.01$\,kpc, $D_{\rm l, max}=10$\,kpc, $\mu_{\rm min}=-200$\,mas\,yr$^{-1}$, $\mu_{\rm max}=200$\,mas\,yr$^{-1}$, $M_{\rm min}=0.01\,M_{\odot}$, and $M_{\rm max}=10\,M_{\odot}$.

\begin{deluxetable}{lcc}
\centering
\tablecaption{Milky Way Disk Models\label{tab:mw_disk}}
\tablehead{
\colhead{} & \colhead{\citealt{han_gould2003}} & \colhead{\citealt{cautun2020}}
}
\startdata
Thin disk \\
$h_R$ (kpc) & 2.75  & 2.63  \\
$h_z$ (kpc) & 0.156 & 0.30 \\
$\rho_0$ ($M_{\odot}\,\mathrm{pc}^{-3}$)   & 0.6623 & 1.2183 \\
\hline
Thick disk\\
$h_R$ (kpc) & 2.75  & 3.80  \\
$h_z$ (kpc) & 0.439 & 0.90 \\
$\rho_0$ ($M_{\odot}\,\mathrm{pc}^{-3}$) & 0.4077 & 0.0561 \\
\enddata
\end{deluxetable}

\subsection{Comparison with Observations}

Table~\ref{tab:num_events} presents the expected number of microlensing events toward the SMC with sources brighter than $I=22$~mag, separately for $\beta=0$ and $\beta=1$. (Recall that we assume that the number of observable sources scales as $\rho(\Ds)\Ds^{2-2\beta}$). For self-lensing models, we find that those with $\beta=0$ predict 6--11\% more events than those with $\beta=1$. In contrast, the expected number of events with lenses located in the Milky Way disk does not significantly depend on the choice of $\beta$. 

Depending on the model selected, we expect 6.7--7.3 self-lensing events and 1.6--3.3 events with Milky Way lenses. This should be compared to the total of five events included in our statistical sample (Section~\ref{sec:search}). For example, the Poisson probability of observing five events when more than 8.3 events are expected is just 0.165. The expected number of events is highly sensitive to the total stellar mass of the SMC, which is not precisely known. If we adopt the mass of $M_* = 4.6 \times 10^8\,M_{\odot}$ \citep{mcconnachie2012}, then the number of expected self-lensing events is scaled down to 3.1--3.4 ($\beta=0$). Then, the total number of expected events is 4.7--6.7, in better agreement with the observed value.

\begin{deluxetable}{lcc}
\centering
\tablecaption{Expected Number of Events\label{tab:num_events}}
\tablehead{
\colhead{Model} & \colhead{$\beta=0$} & \colhead{$\beta=1$}
}
\startdata
SMC self-lensing \citep{calchi_novati2013} & 6.7 & 6.3 \\
SMC self-lensing \citep{mroz_poleski2018}  & 7.3 & 6.6 \\
Milky Way disk \citep{cautun2020}          & 3.3 & 3.3 \\
Milky Way disk \citep{han_gould2003}       & 1.6 & 1.6 \\
\enddata
\end{deluxetable}

\begin{figure*}[htbp]
\includegraphics[width=0.49\textwidth]{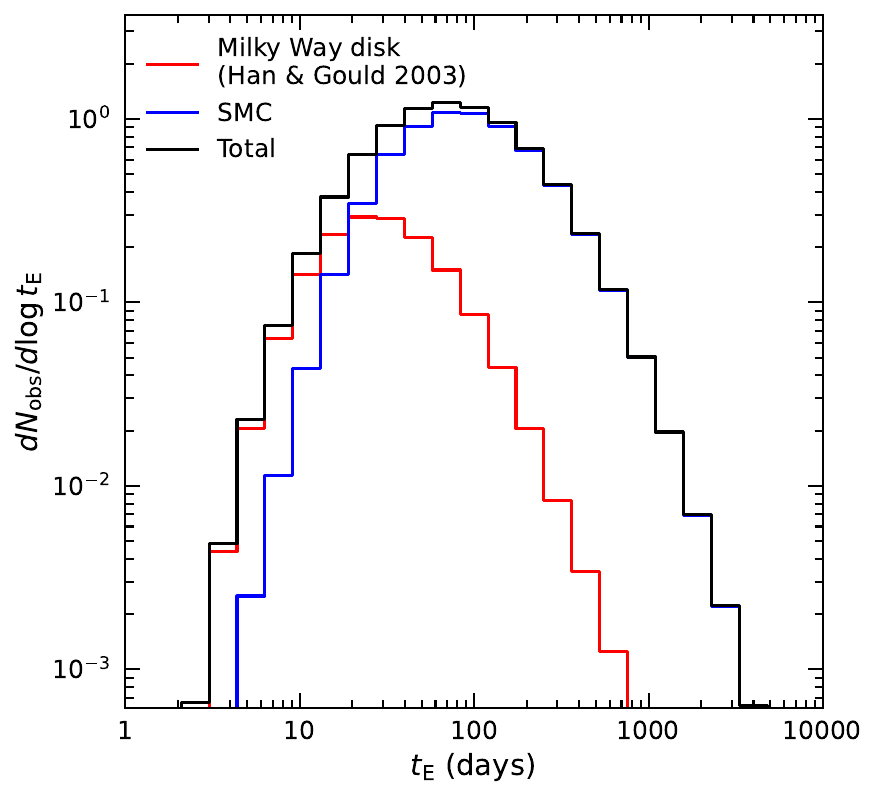}
\includegraphics[width=0.49\textwidth]{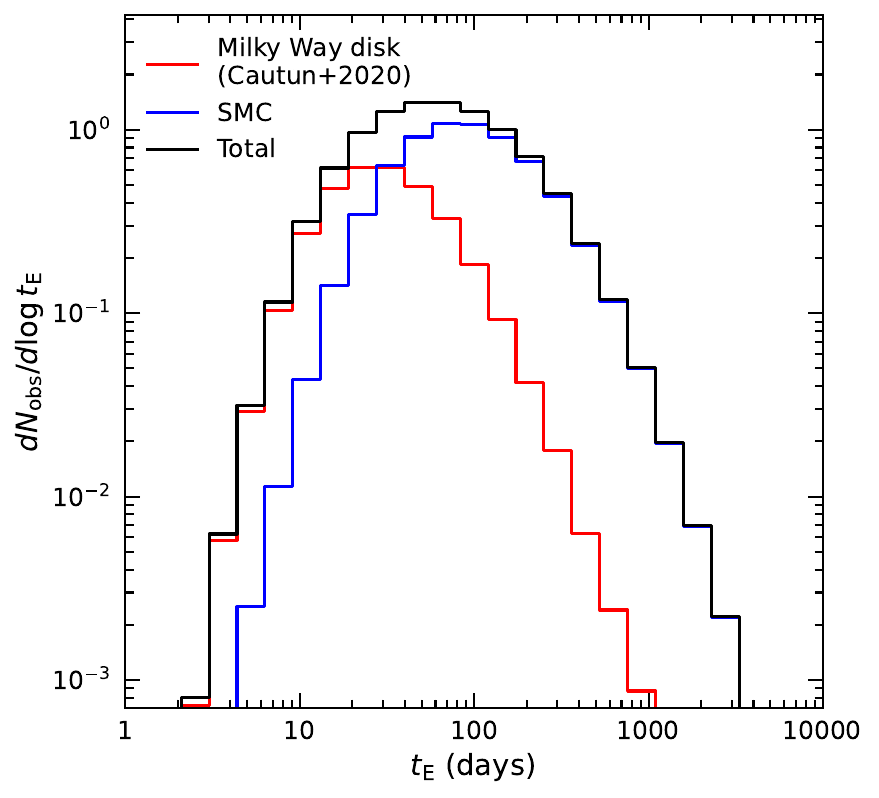}
\caption{Expected distributions of observed event timescales of microlensing events from known stellar populations in the SMC (\citealt{calchi_novati2013}; blue line) and the foreground Milky Way disk (red line). Left panel: assuming the Milky Way model of \citet{han_gould2003}; right panel: \citet{cautun2020}.}
\label{fig:expected_timescales}
\end{figure*}

Figure~\ref{fig:expected_timescales} shows the expected distributions of observed event timescales from known stellar populations in the SMC and the Milky Way disk:
\begin{equation}
\frac{dN_{\rm obs}}{d\log\tE} = \sum_{i} \frac{d\Gamma_i}{d\log\tE}\varepsilon_i(\tE)N_{{\rm s},i} \Delta T_i,
\end{equation}
where the summation is performed over all observed fields, $N_{{\rm s},i}$ is the number of source stars brighter than $I=22$~mag in a given field, and $\Delta T_i$ is the duration of its observations.
Self-lensing events are expected to have relatively long timescales, with the average Einstein timescale being $\langle\tE\rangle \approx 110$\,days, primarily because of low relative lens--source proper motions. In contrast, events with Milky Way lenses tend to have much shorter timescales due to higher relative proper motions, leading to an average of $\langle\tE\rangle \approx 30$\,days.

\begin{figure}
\includegraphics[width=.5\textwidth]{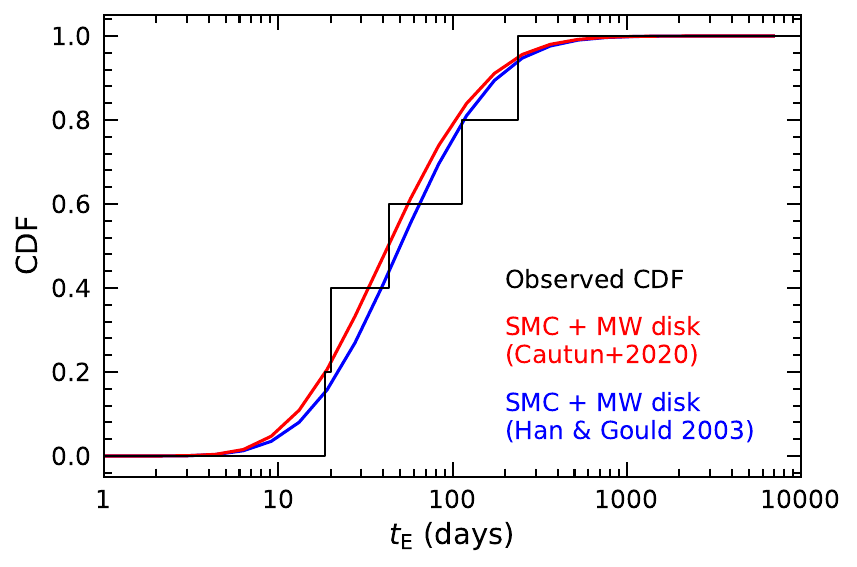}
\caption{Expected cumulative distribution of event timescales of microlensing events from known stellar populations in the SMC and the foreground Milky Way disk, assuming the Milky Way models of \citet{cautun2020} (blue line) and \citet{han_gould2003} (red line). The observed cumulative distribution function is shown by the black line.}
\label{fig:cumulative_timescales}
\end{figure}

The comparison between the observed cumulative distribution of event timescales and the predictions from our models is presented in Figure~\ref{fig:cumulative_timescales}. The expected distributions match the observations reasonably well. To quantify that, we performed a Kolmogorov--Smirnov test, resulting in $p$-values of 0.957 and 0.965 for the models including \citet{han_gould2003} and \citet{cautun2020} Milky Way disk models, respectively (for $\beta=0$ and the \citet{calchi_novati2013} SMC model). We found similar $p$-values for models with $\beta=1$ and those calculated using the \citet{mroz_poleski2018} approach. The results of these tests confirm that there are not any statistically significant deviations between the observed and expected timescales.

\section{Limits on Primordial Black Holes in Dark Matter}
\label{sec:limits}

The results presented in the previous section indicate that all microlensing events detected by OGLE in the SMC direction can be easily explained by the known stellar populations located either within the SMC or in the foreground Milky Way disk. Thus, there is no need to invoke compact objects in the Milky Way dark matter halo to explain them. 

We use the same methodology as described in \citetalias{mroz2024a} to infer upper limits on the abundance of PBHs in dark matter. We quantify this by introducing the parameter $f=M_{\rm PBH}/M_{\rm DM}$, where $M_{\rm DM}$ is the total mass of the dark matter halo (inferred from rotation curve measurements), and $M_{\rm PBH}$ is the total mass of PBHs within it. We note that these limits must apply not only to PBHs but also any other types of compact objects (dark matter subhalos, boson stars, quark nuggets; e.g., \citealt{croon2020}).

The limits are calculated under the assumption of a delta-function mass function of PBH, meaning we assume all PBHs have the identical mass $M$. While some PBH formation models predict more complex and extended PBH mass functions, our results are not highly sensitive to the shape of the mass function as long as most PBHs fall within the mass range probed by microlensing (as discussed in \citetalias{mroz2024a}). In the case of extended mass functions, the limits depend on the choice of the specific model. Therefore, for simplicity, we stick to a delta-function PBH mass function, as we cannot analyze all possible PBH models. Nonetheless, we publish our detection efficiencies and other supporting data, allowing any compelling PBH model that predicts an extended mass function to be independently tested.

We assume that the distribution of PBHs within the dark matter halo can be described by the contracted halo model of \citet{cautun2020}. This model was fitted to the Milky Way's rotation curve, which was inferred from the Gaia DR2 data \citep{eilers2019} and other sources. For the PBHs velocities, we use a Maxwellian distribution with the standard deviation of velocities in one direction set to $V_0(R)/\sqrt{2}$, where $V_0(R)$ is the Milky Way rotation curve predicted by the model. We neglect dark matter located in the SMC halo, as its mass (about $6.5 \times 10^9\,M_{\odot}$; \citealt{bekki2009b}) is significantly smaller than that of the Milky Way halo. The expected number of microlensing events in a given field was calculated using Equations~(\ref{eq:gamma}) and~(\ref{eq:n_exp}), assuming $\beta=0$ and employing the model of \citet{calchi_novati2013} for the distribution of source stars.

\begin{figure}
\includegraphics[width=.5\textwidth]{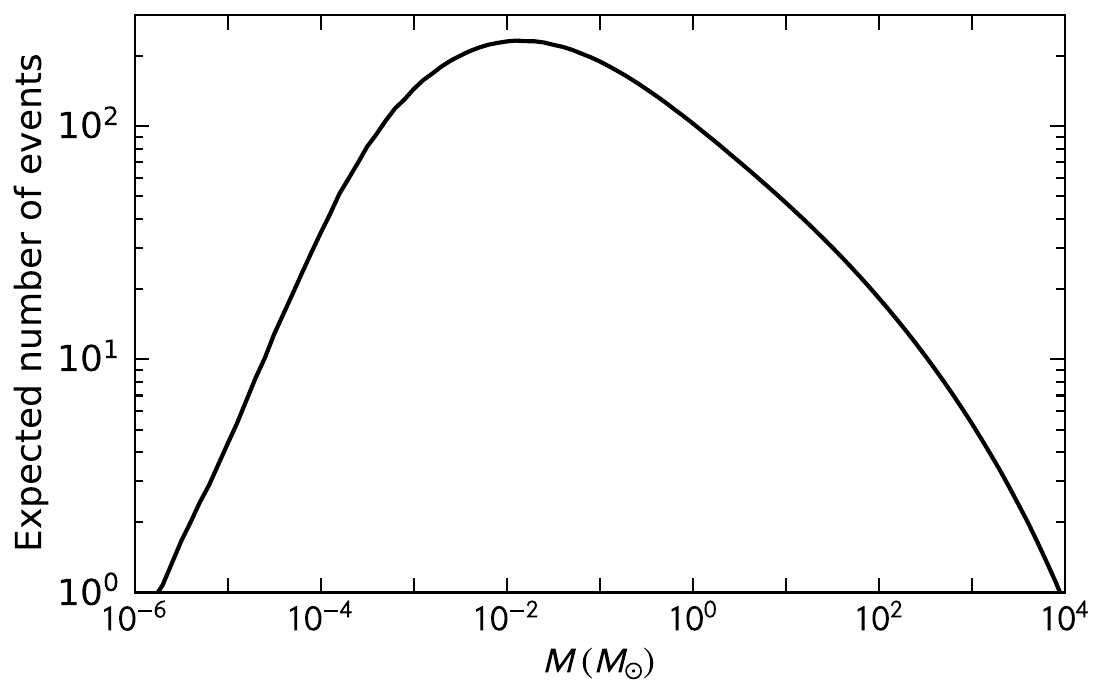}
\caption{Expected number of microlensing events to be detected in the SMC direction if the entire Milky Way's dark matter halo was composed of compact objects with identical mass.}
\label{fig:n_expected}
\end{figure}

Figure~\ref{fig:n_expected} shows the total expected number of microlensing events, assuming that the entire Milky Way dark matter halo consists of compact objects of identical mass $M$. We expect the largest number of events (approximately 233) for $M=0.013\,M_{\odot}$. We should have detected about 103 events if the Milky Way's entire dark matter halo was made of $1\,M_{\odot}$ PBHs, 47 events---$10\,M_{\odot}$, and 18 events---$100\,M_{\odot}$. That should be compared to five events in our statistical sample (Section~\ref{sec:search}).

\begin{figure*}
\centering
\includegraphics[width=.7\textwidth]{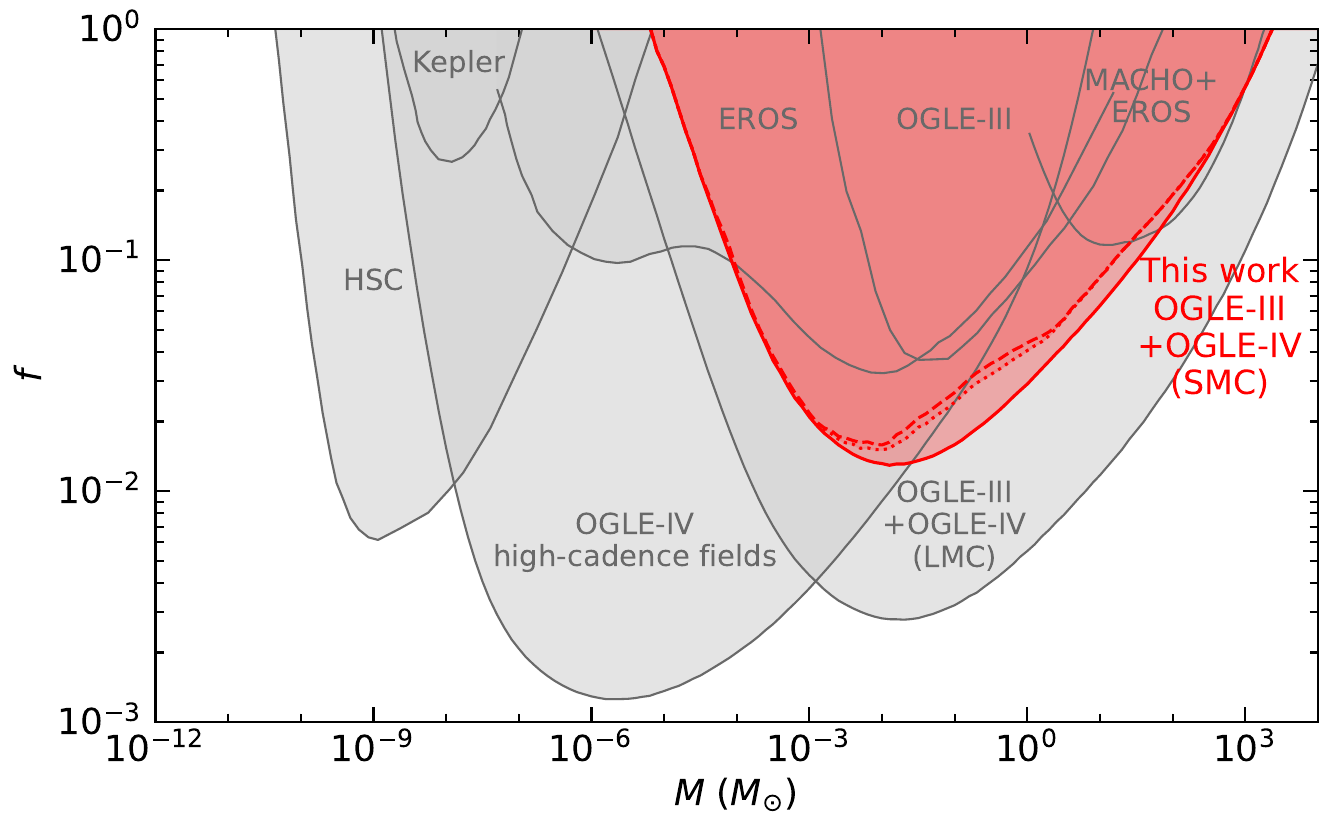}
\caption{95\% upper limits on the fraction of dark matter in the form of PBHs and other compact objects. The shaded red area is excluded at 95\% confidence based on the presented observations of the SMC. The solid red line represents the strict limits, whereas the dashed and dotted red lines indicate the relaxed limits for the \citet{han_gould2003} and \citet{cautun2020} Milky Way disk models, respectively. Other limits are taken from \citet{tisserand2007} (EROS), \citet{wyrzyk1} (\mbox{OGLE-III}), \citet{griest2013} (Kepler), \citet{niikura2019} (HSC), \citet{moniez2022} (EROS+MACHO), \citet{mroz2024a} (\mbox{OGLE-III} + \mbox{OGLE-IV}), and \citet{mroz2024c} (\mbox{OGLE-IV} high-cadence).}
\label{fig:bounds}
\end{figure*}

We calculated the 95\% upper limits on $f$ using the same methodology as in \citetalias{mroz2024a}. We present two types of limits. For the strict limits, we assumed that all five microlensing events in our statistical sample are due to known stellar populations in the SMC or in the Milky Way disk. These limits are marked with a solid red line in Figure~\ref{fig:bounds}. When we relax this assumption and allow for a possibility of a PBH origin for these microlensing events, the inferred limits become slightly weaker. They are shown by red dashed and dotted lines in Figure~\ref{fig:bounds}. The relaxed limits slightly depend on the choice of the Milky Way disk or the SMC model. In Figure~\ref{fig:bounds} we present two example results for the \citet{han_gould2003} (dashed line) and \citet{cautun2020} (dotted line) Milky Way disk models.

The strict limits reach $f=1.3\%$ for $M=0.013\,M_{\odot}$ for which we expect the largest number of microlensing events. We find that less than $f=2.9\%$ of dark matter may be composed of compact objects of $1\,M_{\odot}$, less than $f=6.4\%$ for $10\,M_{\odot}$, less than $f=16\%$ for $100\,M_{\odot}$. The difference between strict and relaxed limits (for the \citet{han_gould2003} Milky Way disk model) is largest for $\log M=-0.7$ and amounts to 0.24\,dex (that is, the relaxed limits are a factor of 1.8 weaker for that mass). 

\section{Discussion and Conclusions}
\label{sec:dis}

\subsection{Optical Depth toward the SMC}

The only previously published measurements of the microlensing optical depth toward the SMC were by \citet{wyrzyk3} and \citet{wyrzyk1}, based on the \mbox{OGLE-II} and \mbox{OGLE-III} data, respectively. \citet{wyrzyk3} measured the optical depth of $\tau=(1.15 \pm 1.15) \times 10^{-7}$ based on just one event candidate, OGLE-SMC-01. However, as we found that this object is a variable star rather than a microlensing event, this measurement should be regarded as an upper limit. \citet{wyrzyk1} found $\tau = (1.30 \pm 1.01) \times 10^{-7}$ based on a sample of three events. Of these three events, OGLE-SMC-02 had the largest contribution to this measurement (about 65\%). The corresponding microlensing event rate is $\Gamma = (4.4 \pm 2.6) \times 10^{-7}\,\mathrm{yr}^{-1}\,\mathrm{star}^{-1}$.

Using the combined \mbox{OGLE-III} and \mbox{OGLE-IV} data set, we found smaller values for both the optical depth ($\tau = (0.32 \pm 0.18) \times 10^{-7}$) and the event rate ($\Gamma = (1.18 \pm 0.57) \times 10^{-7}\,\mathrm{yr}^{-1}\,\mathrm{star}^{-1}$). However, because of the small sample sizes, these measurements are formally statistically consistent. For the same reason, the optical depth measurements are prone to large statistical fluctuations. For example, including OGLE-SMC-02 in our statistical sample would increase the optical depth by 20\%. 

One difference between our work and that of \citet{wyrzyk1} is that we averaged the optical depth over a much larger area than that observed by \mbox{OGLE-III} and used sources brighter than $I=22$ mag. In contrast, \citet{wyrzyk1} took into account sources brighter than $I=21$ mag. When we restricted our measurements to the \mbox{OGLE-III} footprint and used only sources brighter than $I=21$ mag, we found $\tau = (0.52 \pm 0.29) \times 10^{-7}$ and $\Gamma = (1.76 \pm 0.83) \times 10^{-7}\,\mathrm{yr}^{-1}\,\mathrm{star}^{-1}$, in better agreement with the findings of \citet{wyrzyk1}. 

\subsection{Limits on Compact Objects as Dark Matter}

The limits on the abundance of compact objects in dark matter, as derived in the previous section, were calculated for source stars brighter than $I=22$ mag. However, the event detection efficiency substantially drops with the decreasing source brightness, while the uncertainties on the number of source stars become larger. These factors may increase the risk of introducing systematic errors into the analysis. 

\begin{figure}
\includegraphics[width=.5\textwidth]{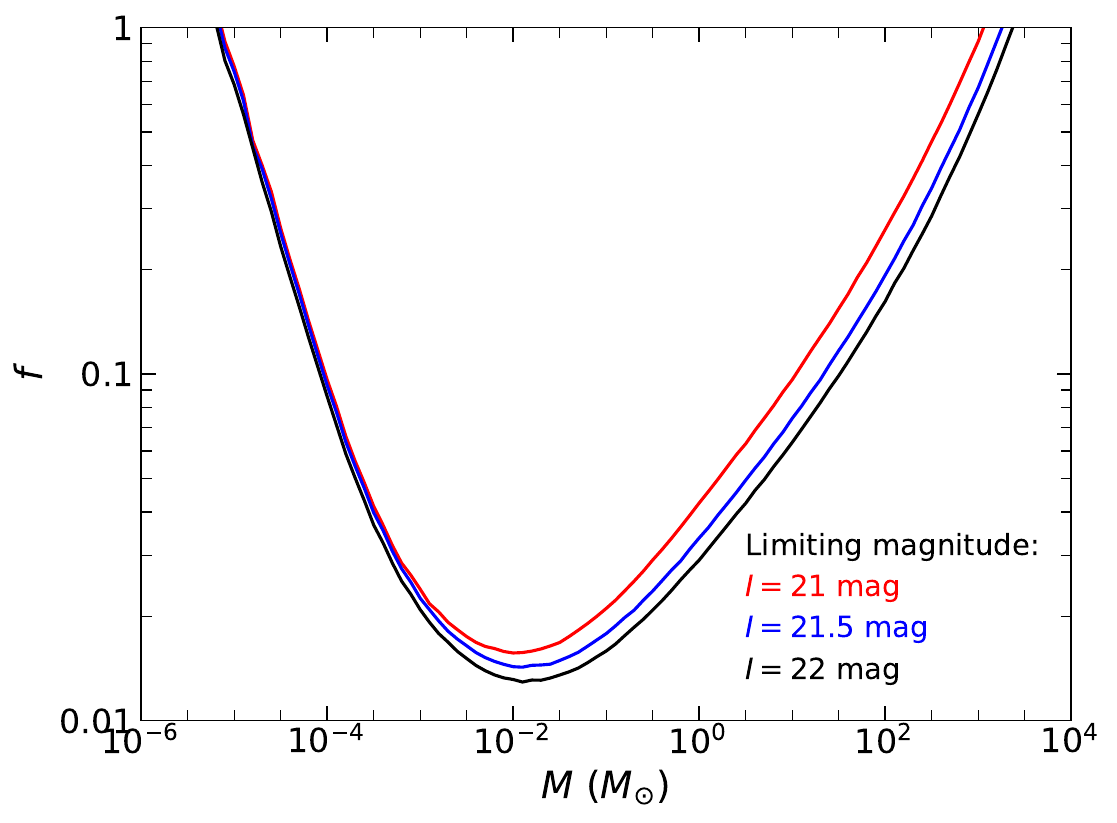}
\caption{Dependence of inferred upper limits on the limiting magnitude.}
\label{fig:bounds_mag}
\end{figure}

To evaluate the possible impact of these systematic errors, we repeated the entire analysis using brighter sources, for which these effects are substantially smaller. The results are presented in Figure~\ref{fig:bounds_mag}, which shows the 95\% upper limits on $f$ derived from sources brighter than $I=21$ (red line), $I=21.5$ (blue line), and $I=22$ mag (black line). As expected, the limits become slightly weaker for brighter sources. However, the differences are minimal; for example, the maximum difference between the limits on $f$ obtained using sources brighter than $I=21$ and $I=22$ mag is only 0.21\,dex.

The inferred limits depend on the choice of the Milky Way halo model. In this study, we used the contracted halo model from \citet{cautun2020}, which was fitted to the Milky Way's rotation curve derived from the Gaia DR2 data \citep{eilers2019}. However, several recent works \citep[e.g.,][]{wang2023,jiao2023,ou2024} have indicated evidence for a declining Milky Way rotation curve at Galactocentric distances greater than 15--20\,kpc, based on the more recent Gaia DR3 data. This suggests that the estimated mass of the Milky Way halo may be smaller than previously thought. These new findings, however, are in tension with other mass estimates derived from observations of globular clusters, dwarf satellites, and streams. \citet{koop2024} pointed out that---at large Galactocentric radii---disequilibrium and deviations from axisymmetry could lead to large systematic errors in the inferred rotation curves.

Nevertheless, we found that our results are not strongly influenced by the choice of these models. The inferred limits on the frequency of PBHs are inversely proportional to the expected number of microlensing events, which in turn is directly proportional to the optical depth in a given direction. For example, the microlensing optical depth due to the Milky Way's dark matter halo in the direction of the SMC center ($l=303^{\circ}$, $b=-44^{\circ}$, $D_{\rm SMC}=62.44$ kpc) in our fiducial model \citep{cautun2020} is $\tau=6.4 \times 10^{-7}$. For the B2 halo model from \citet{jiao2023}, we find $\tau=6.2 \times 10^{-7}$, while the halo model from \citet{ou2024} yields $\tau=6.0 \times 10^{-7}$. These results indicate that the inferred limits on $f$ would be reduced by less than $6\%$ if the halo model from \citet{cautun2020} was replaced by that of \citet{jiao2023} or \citet{ou2024}.

Additionally, \citet{green2025} investigated how the choice of the halo model impacts microlensing constraints on the abundance of PBHs. They considered two power-law halo models \citep{evans1994} from \citet{alcock1996} (models B and C) and found that the tightest limit on $f$ is a factor of 1.7 (1.5) lower (higher) for model B (C) compared to the standard halo model. We note, however, that none of these models accurately fit the Milky Way rotation curve \citep{mroz2019c}: model B consistently predicts higher rotation speeds, while model C predicts lower speeds than actually observed. In contrast, the rotation curves predicted by the models from \citet{cautun2020}, \citet{jiao2023}, and \citet{ou2024} match the data very well. We conclude that the uncertainties on $f$ associated with the choice of the Milky Way halo model are substantially smaller than those indicated by the predictions of the B and C halo models of \citet{alcock1996}.

Recently, \citet{li2025} studied the constraints on the abundance of planetary-mass PBHs based on a sample of 2617 events detected by OGLE in the direction of the Galactic bulge \citep{mroz2017}. They investigated the dependence of these constraints on the choice of dark matter halo models and found that the constraints derived assuming the Einasto and Burkert profiles are much weaker than those from the standard Navarro--Frenk--White profile. We check the same models to study the robustness of our results.

The Einasto profile used by \citet{li2025} was taken from \citet{hertzberg2021}, who in turn used the dark matter profile from \citet{pieri2011}. The latter model was found from the Milky Way \textit{Aquarius} simulations by \citet{springel2008}. Unfortunately, the parameters of the model changed during this telephone game: \citet{pieri2011} reported the normalization of the profile of $\rho_s = 2.8 \times 10^6\,M_{\odot}\,\mathrm{kpc}^{-1}$, while \citet{li2025} used $\rho_s = 2.08 \times 10^6\,M_{\odot}\,\mathrm{kpc}^{-1}$ (thus, the constrains published by \citet{li2025} are underestimated by 35\%). Assuming that the parameters from \citet{pieri2011} are correct, the microlensing optical depth toward the SMC center is $\tau=11.1\times 10^{-7}$, i.e., a factor 1.7 larger than in our fiducial model. However, the Milky Way rotation curve predicted by the former model does not match the data \citep{mroz2019c}. We also found that the rotation curves calculated using Burkert profiles considered in \citet{li2025} do not match the observed one at all \citep{mroz2019c}. That is not surprising, as these models come from \citet{donato2009} and they were intended to fit the weak gravitational lensing observations of some galaxies, but not the Milky Way dark matter profile.

\subsection{Summary}

This paper concludes the current searches for microlensing events toward the Magellanic Clouds using data from the OGLE survey. In \citetalias{mroz2024b} and \citetalias{mroz2024a}, we presented the results of the searches for very long-timescale microlensing events toward the LMC using the combined \mbox{OGLE-III} and \mbox{OGLE-IV} data. Such long-duration events were expected if dark matter in the Milky Way halo consisted of massive PBHs, with masses ranging from dozens to hundreds $M_{\odot}$, similar to black holes detected by the gravitational-wave observatories. In \citetalias{mroz2024c}, we searched for short-duration microlensing events that could have been caused by planetary-mass PBHs. No such events were discovered, which provided very strict limits on the abundance of PBHs and other compact objects in the Milky Way halo with masses ranging from $10^{-8}$ to $10^3\,M_{\odot}$.

In this paper, we extended our search for long-timescale microlensing events to stars located in the SMC. Thanks to combining \mbox{OGLE-III} and \mbox{OGLE-IV} data, we created nearly 20-year-long light curves for about 7.4~million stars. For an additional 6.2~million stars, we used 10-year-long \mbox{OGLE-IV} light curves. The analysis of this data set allowed us to achieve high sensitivity to long-timescale microlensing events. 

We recovered three previously known events and discovered three new ones. We measured the microlensing optical depth toward the SMC to be $\tau = (0.32 \pm 0.18) \times 10^{-7}$ and the event rate to be $\Gamma = (1.18 \pm 0.57) \times 10^{-7}\,\mathrm{yr}^{-1}\,\mathrm{star}^{-1}$. The properties of the detected events are consistent with lenses originating from known stellar populations within the SMC or in the Milky Way disk, leaving little room for PBHs or other dark compact objects in the dark matter halo.

In particular, our findings indicate that PBHs with a mass of $1\,M_{\odot}$ may account for at most $2.9\%$ of the dark matter in the Milky Way, $10\,M_{\odot}$---6.4\%, $100\,M_{\odot}$---16\%. While these limits are weaker than those derived in \citetalias{mroz2024a}, they are still stronger than the results reported by \citet{tisserand2007}, \citet{wyrzyk1}, and \citet{moniez2022} for the mass range of $10^{-4}\!-\!10^2 M_{\odot}$. In addition, our results provide independent information about the number of compact objects along a different line of sight through the Galactic halo.

The data presented in this paper are publicly available at \url{https://www.astrouw.edu.pl/ogle/ogle4/SMC_OPTICAL_DEPTH/} and via doi: \dataset[105281/zenodo.16735482]{\doi{105281/zenodo.16735482}}.

\section*{Acknowledgements}

We thank the anonymous referee for providing constructive comments that helped improve the paper.
We thank all of the OGLE observers for their contribution to the collection of the photometric data over the decades. We thank \L{}ukasz Wyrzykowski for his comments on the manuscript. This research was funded in part by National Science Centre, Poland, grants OPUS 2021/41/B/ST9/00252 and SONATA 2023/51/D/ST9/00187 awarded to P.M. The OGLE project has received funding from the Polish National Science Centre grant OPUS 2024/55/B/ST9/00447 awarded to A.U.

\bibliographystyle{aasjournal}
\bibliography{pap}

\begin{thebibliography}{}
\expandafter\ifx\csname natexlab\endcsname\relax\def\natexlab#1{#1}\fi

\bibitem[{{Afonso} {et~al.}(2000){Afonso}, {Alard}, {Albert}, {Andersen},
  {Ansari}, {Aubourg}, {Bareyre}, {Bauer}, {Beaulieu}, {Bouquet}, {Char},
  {Charlot}, {Couchot}, {Coutures}, {Derue}, {Ferlet}, {Glicenstein},
  {Goldman}, {Gould}, {Graff}, {Gros}, {Haissinski}, {Hamilton}, {Hardin}, {de
  Kat}, {Kim}, {Lasserre}, {Lesquoy}, {Loup}, {Magneville}, {Marquette},
  {Maurice}, {Milsztajn}, {Moniez}, {Palanque-Delabrouille}, {Perdereau},
  {Pr{\'e}vot}, {Regnault}, {Rich}, {Spiro}, {Vidal-Madjar}, {Vigroux},
  {Zylberajch}, {Alcock}, {Allsman}, {Alves}, {Axelrod}, {Becker}, {Cook},
  {Drake}, {Freeman}, {Griest}, {King}, {Lehner}, {Marshall}, {Minniti},
  {Peterson}, {Pratt}, {Quinn}, {Rodgers}, {Stetson}, {Stubbs}, {Sutherland},
  {Tomaney}, {Vandehei}, {Rhie}, {Bennett}, {Fragile}, {Johnson}, {Quinn},
  {Udalski}, {Kubiak}, {Szyma{\'n}ski}, {Pietrzy{\'n}ski}, {Wo{\'z}niak},
  {Zebru{\'n}}, {Albrow}, {Caldwell}, {DePoy}, {Dominik}, {Gaudi}, {Greenhill},
  {Hill}, {Kane}, {Martin}, {Menzies}, {Naber}, {Pogge}, {Pollard}, {Sackett},
  {Sahu}, {Vermaak}, {Watson}, \& {Williams}}]{afonso2000}
{Afonso}, C., {Alard}, C., {Albert}, J.~N., {et~al.} 2000, \apj, 532, 340

\bibitem[{{Alard} \& {Lupton}(1998)}]{alard1998}
{Alard}, C., \& {Lupton}, R.~H. 1998, \apj, 503, 325

\bibitem[{{Alcock} {et~al.}(1996){Alcock}, {Allsman}, {Axelrod}, {Bennett},
  {Cook}, {Freeman}, {Griest}, {Guern}, {Lehner}, {Marshall}, {Park},
  {Perlmutter}, {Peterson}, {Pratt}, {Quinn}, {Rodgers}, {Stubbs}, \&
  {Sutherland}}]{alcock1996}
{Alcock}, C., {Allsman}, R.~A., {Axelrod}, T.~S., {et~al.} 1996, \apj, 461, 84

\bibitem[{{Alcock} {et~al.}(1997{\natexlab{a}}){Alcock}, {Allsman}, {Alves},
  {Axelrod}, {Becker}, {Bennett}, {Cook}, {Freeman}, {Griest}, {Keane},
  {Lehner}, {Marshall}, {Minniti}, {Peterson}, {Pratt}, {Quinn}, {Rodgers},
  {Stubbs}, {Sutherland}, {Tomaney}, {Vandehei}, \& {Welch}}]{alcock1997_smc}
{Alcock}, C., {Allsman}, R.~A., {Alves}, D., {et~al.} 1997{\natexlab{a}},
  \apjl, 491, L11

\bibitem[{{Alcock} {et~al.}(1997{\natexlab{b}}){Alcock}, {Allsman}, {Alves},
  {Axelrod}, {Becker}, {Bennett}, {Cook}, {Freeman}, {Griest}, {Guern},
  {Lehner}, {Marshall}, {Peterson}, {Pratt}, {Quinn}, {Rodgers}, {Stubbs},
  {Sutherland}, \& {Welch}}]{alcock1997c}
{Alcock}, C., {Allsman}, R.~A., {Alves}, D., {et~al.} 1997{\natexlab{b}}, \apj,
  486, 697

\bibitem[{{Alcock} {et~al.}(1999){Alcock}, {Allsman}, {Alves}, {Axelrod},
  {Becker}, {Bennett}, {Cook}, {Drake}, {Freeman}, {Griest}, {King}, {Lehner},
  {Marshall}, {Minniti}, {Peterson}, {Pratt}, {Quinn}, {Rhie}, {Rodgers},
  {Stetson}, {Stubbs}, {Sutherland}, {Tomaney}, \& {Vandehei}}]{alcock1999}
{Alcock}, C., {Allsman}, R.~A., {Alves}, D., {et~al.} 1999, \apj, 518, 44

\bibitem[{{Alcock} {et~al.}(2000){Alcock}, {Allsman}, {Alves}, {Axelrod},
  {Becker}, {Bennett}, {Cook}, {Dalal}, {Drake}, {Freeman}, {Geha}, {Griest},
  {Lehner}, {Marshall}, {Minniti}, {Nelson}, {Peterson}, {Popowski}, {Pratt},
  {Quinn}, {Stubbs}, {Sutherland}, {Tomaney}, {Vandehei}, \&
  {Welch}}]{alcock2000b}
{Alcock}, C., {Allsman}, R.~A., {Alves}, D.~R., {et~al.} 2000, \apj, 542, 281

\bibitem[{{Batista} {et~al.}(2011){Batista}, {Gould}, {Dieters}, {Dong},
  {Bond}, {Beaulieu}, {Maoz}, {Monard}, {Christie}, {McCormick}, {Albrow},
  {Horne}, {Tsapras}, {Burgdorf}, {Calchi Novati}, {Skottfelt}, {Caldwell},
  {Koz{\l}owski}, {Kubas}, {Gaudi}, {Han}, {Bennett}, {An}, {MOA
  Collaboration}, {Abe}, {Botzler}, {Douchin}, {Freeman}, {Fukui}, {Furusawa},
  {Hearnshaw}, {Hosaka}, {Itow}, {Kamiya}, {Kilmartin}, {Korpela}, {Lin},
  {Ling}, {Makita}, {Masuda}, {Matsubara}, {Miyake}, {Muraki}, {Nagaya},
  {Nishimoto}, {Ohnishi}, {Okumura}, {Perrott}, {Rattenbury}, {Saito},
  {Sullivan}, {Sumi}, {Sweatman}, {Tristram}, {von Seggern}, {Yock}, {PLANET
  Collaboration}, {Brillant}, {Calitz}, {Cassan}, {Cole}, {Cook}, {Coutures},
  {Dominis Prester}, {Donatowicz}, {Greenhill}, {Hoffman}, {Jablonski}, {Kane},
  {Kains}, {Marquette}, {Martin}, {Martioli}, {Meintjes}, {Menzies},
  {Pedretti}, {Pollard}, {Sahu}, {Vinter}, {Wambsganss}, {Watson}, {Williams},
  {Zub}, {FUN Collaboration}, {Allen}, {Bolt}, {Bos}, {DePoy}, {Drummond},
  {Eastman}, {Gal-Yam}, {Gorbikov}, {Higgins}, {Janczak}, {Kaspi}, {Lee},
  {Mallia}, {Maury}, {Monard}, {Moorhouse}, {Morgan}, {Natusch}, {Ofek},
  {Park}, {Pogge}, {Polishook}, {Santallo}, {Shporer}, {Spector}, {Thornley},
  {Yee}, {MiNDSTEp Consortium}, {Bozza}, {Browne}, {Dominik}, {Dreizler},
  {Finet}, {Glitrup}, {Grundahl}, {Harps{\o}e}, {Hessman}, {Hinse},
  {Hundertmark}, {J{\o}rgensen}, {Liebig}, {Maier}, {Mancini}, {Mathiasen},
  {Rahvar}, {Ricci}, {Scarpetta}, {Southworth}, {Surdej}, {Zimmer}, {RoboNet
  Collaboration}, {Allan}, {Bramich}, {Snodgrass}, {Steele}, \&
  {Street}}]{batista2011}
{Batista}, V., {Gould}, A., {Dieters}, S., {et~al.} 2011, \aap, 529, A102

\bibitem[{{Bechtol} {et~al.}(2022){Bechtol}, {Birrer}, {Cyr-Racine}, {Schutz},
  {Adhikari}, {Amin}, {Banerjee}, {Bird}, {Blinov}, {Boddy}, {Boehm}, {Bundy},
  {Buschmann}, {Chakrabarti}, {Curtin}, {Dai}, {Drlica-Wagner}, {Dvorkin},
  {Erickcek}, {Gilman}, {Heeba}, {Kim}, {Ir{\v{s}}i{\v{c}}}, {Leauthaud},
  {Lovell}, {Luki{\'c}}, {Mao}, {Mau}, {Mitridate}, {Mocz}, {Mu{\~n}oz},
  {Nadler}, {Peter}, {Price-Whelan}, {Robertson}, {Sabti}, {Sehgal}, {Shipp},
  {Simon}, {Singh}, {Van Tilburg}, {Wechsler}, {Widmark}, \&
  {Yu}}]{bechtol2022}
{Bechtol}, K., {Birrer}, S., {Cyr-Racine}, F.-Y., {et~al.} 2022, arXiv
  e-prints, arXiv:2203.07354

\bibitem[{{Bekki} \& {Chiba}(2009)}]{bekki2009}
{Bekki}, K., \& {Chiba}, M. 2009, \pasa, 26, 37

\bibitem[{{Bekki} \& {Stanimirovi{\'c}}(2009)}]{bekki2009b}
{Bekki}, K., \& {Stanimirovi{\'c}}, S. 2009, \mnras, 395, 342

\bibitem[{{Bird} {et~al.}(2016){Bird}, {Cholis}, {Mu{\~n}oz},
  {Ali-Ha{\"\i}moud}, {Kamionkowski}, {Kovetz}, {Raccanelli}, \&
  {Riess}}]{bird2016}
{Bird}, S., {Cholis}, I., {Mu{\~n}oz}, J.~B., {et~al.} 2016, \prl, 116, 201301

\bibitem[{{Blaineau} {et~al.}(2022){Blaineau}, {Moniez}, {Afonso}, {Albert},
  {Ansari}, {Aubourg}, {Coutures}, {Glicenstein}, {Goldman}, {Hamadache},
  {Lasserre}, {Le Guillou}, {Lesquoy}, {Magneville}, {Marquette},
  {Palanque-Delabrouille}, {Perdereau}, {Rich}, {Spiro}, \&
  {Tisserand}}]{moniez2022}
{Blaineau}, T., {Moniez}, M., {Afonso}, C., {et~al.} 2022, \aap, 664, A106

\bibitem[{{Calchi Novati}(2010)}]{calchi_novati2010}
{Calchi Novati}, S. 2010, General Relativity and Gravitation, 42, 2101

\bibitem[{{Calchi Novati} {et~al.}(2013){Calchi Novati}, {Mirzoyan}, {Jetzer},
  \& {Scarpetta}}]{calchi_novati2013}
{Calchi Novati}, S., {Mirzoyan}, S., {Jetzer}, P., \& {Scarpetta}, G. 2013,
  \mnras, 435, 1582

\bibitem[{{Calchi Novati} {et~al.}(2005){Calchi Novati}, {Paulin-Henriksson},
  {An}, {Baillon}, {Belokurov}, {Carr}, {Cr{\'e}z{\'e}}, {Evans},
  {Giraud-H{\'e}raud}, {Gould}, {Hewett}, {Jetzer}, {Kaplan}, {Kerins},
  {Smartt}, {Stalin}, {Tsapras}, {Weston}, \& {POINT-AGAPE
  Collaboration}}]{calchi_novati2005}
{Calchi Novati}, S., {Paulin-Henriksson}, S., {An}, J., {et~al.} 2005, \aap,
  443, 911

\bibitem[{{Carr} {et~al.}(2021){Carr}, {Kohri}, {Sendouda}, \&
  {Yokoyama}}]{carr2021}
{Carr}, B., {Kohri}, K., {Sendouda}, Y., \& {Yokoyama}, J. 2021, Reports on
  Progress in Physics, 84, 116902

\bibitem[{{Carr} {et~al.}(2024){Carr}, {Clesse}, {Garc{\'\i}a-Bellido},
  {Hawkins}, \& {K{\"u}hnel}}]{carr2024}
{Carr}, B.~J., {Clesse}, S., {Garc{\'\i}a-Bellido}, J., {Hawkins}, M.~R.~S., \&
  {K{\"u}hnel}, F. 2024, \physrep, 1054, 1

\bibitem[{{Carr} \& {Hawking}(1974)}]{carr1974}
{Carr}, B.~J., \& {Hawking}, S.~W. 1974, \mnras, 168, 399

\bibitem[{{Cautun} {et~al.}(2020){Cautun}, {Ben{\'\i}tez-Llambay}, {Deason},
  {Frenk}, {Fattahi}, {G{\'o}mez}, {Grand}, {Oman}, {Navarro}, \&
  {Simpson}}]{cautun2020}
{Cautun}, M., {Ben{\'\i}tez-Llambay}, A., {Deason}, A.~J., {et~al.} 2020,
  \mnras, 494, 4291

\bibitem[{{Chapline}(1975)}]{chapline1975}
{Chapline}, G.~F. 1975, \nat, 253, 251

\bibitem[{{Clesse} \& {Garc{\'\i}a-Bellido}(2017)}]{clesse2017}
{Clesse}, S., \& {Garc{\'\i}a-Bellido}, J. 2017, Physics of the Dark Universe,
  15, 142

\bibitem[{{Croon} {et~al.}(2020){Croon}, {McKeen}, \& {Raj}}]{croon2020}
{Croon}, D., {McKeen}, D., \& {Raj}, N. 2020, \prd, 101, 083013

\bibitem[{{de Jong} {et~al.}(2006){de Jong}, {Widrow}, {Cseresnjes}, {Kuijken},
  {Crotts}, {Bergier}, {Baltz}, {Gyuk}, {Sackett}, {Uglesich}, \&
  {Sutherland}}]{de_jong2006}
{de Jong}, J.~T.~A., {Widrow}, L.~M., {Cseresnjes}, P., {et~al.} 2006, \aap,
  446, 855

\bibitem[{{Despali} {et~al.}(2018){Despali}, {Vegetti}, {White}, {Giocoli}, \&
  {van den Bosch}}]{despali2018}
{Despali}, G., {Vegetti}, S., {White}, S. D.~M., {Giocoli}, C., \& {van den
  Bosch}, F.~C. 2018, \mnras, 475, 5424

\bibitem[{{Donato} {et~al.}(2009){Donato}, {Gentile}, {Salucci}, {Frigerio
  Martins}, {Wilkinson}, {Gilmore}, {Grebel}, {Koch}, \& {Wyse}}]{donato2009}
{Donato}, F., {Gentile}, G., {Salucci}, P., {et~al.} 2009, \mnras, 397, 1169

\bibitem[{{Dong} {et~al.}(2007){Dong}, {Udalski}, {Gould}, {Reach}, {Christie},
  {Boden}, {Bennett}, {Fazio}, {Griest}, {Szyma{\'n}ski}, {Kubiak},
  {Soszy{\'n}ski}, {Pietrzy{\'n}ski}, {Szewczyk}, {Wyrzykowski}, {Ulaczyk},
  {Wieckowski}, {Paczy{\'n}ski}, {DePoy}, {Pogge}, {Preston}, {Thompson}, \&
  {Patten}}]{dong2007}
{Dong}, S., {Udalski}, A., {Gould}, A., {et~al.} 2007, \apj, 664, 862

\bibitem[{{Eilers} {et~al.}(2019){Eilers}, {Hogg}, {Rix}, \&
  {Ness}}]{eilers2019}
{Eilers}, A.-C., {Hogg}, D.~W., {Rix}, H.-W., \& {Ness}, M.~K. 2019, \apj, 871,
  120

\bibitem[{{Evans}(1994)}]{evans1994}
{Evans}, N.~W. 1994, \mnras, 267, 333

\bibitem[{{Gaia Collaboration} {et~al.}(2016){Gaia Collaboration}, {Prusti},
  {de Bruijne}, {Brown}, {Vallenari}, {Babusiaux}, {Bailer-Jones}, {Bastian},
  {Biermann}, {Evans}, \& et~al.}]{gaia2016}
{Gaia Collaboration}, {Prusti}, T., {de Bruijne}, J.~H.~J., {et~al.} 2016,
  \aap, 595, A1

\bibitem[{{Gaia Collaboration} {et~al.}(2018){Gaia Collaboration}, {Helmi},
  {van Leeuwen}, {McMillan}, {Massari}, {Antoja}, {Robin}, {Lindegren},
  {Bastian}, {Arenou}, {Babusiaux}, {Biermann}, {Breddels}, {Hobbs}, {Jordi},
  {Pancino}, {Reyl{\'e}}, {Veljanoski}, {Brown}, {Vallenari}, {Prusti}, {de
  Bruijne}, {Bailer-Jones}, {Evans}, {Eyer}, {Jansen}, {Klioner}, {Lammers},
  {Luri}, {Mignard}, {Panem}, {Pourbaix}, {Randich}, {Sartoretti}, {Siddiqui},
  {Soubiran}, {Walton}, {Cropper}, {Drimmel}, {Katz}, {Lattanzi}, {Bakker},
  {Cacciari}, {Casta{\~n}eda}, {Chaoul}, {Cheek}, {De Angeli}, \&
  {Fabricius}}]{helmi2018}
{Gaia Collaboration}, {Helmi}, A., {van Leeuwen}, F., {et~al.} 2018, \aap, 616,
  A12

\bibitem[{{Gaia Collaboration} {et~al.}(2021{\natexlab{a}}){Gaia
  Collaboration}, {Luri}, {Chemin}, {Clementini}, {Delgado}, {McMillan},
  {Romero-G{\'o}mez}, {Balbinot}, {Castro-Ginard}, {Mor}, {Ripepi}, {Sarro},
  {Cioni}, {Fabricius}, {Garofalo}, {Helmi}, {Muraveva}, {Brown}, {Vallenari},
  {Prusti}, {de Bruijne}, {Babusiaux}, {Biermann}, {Creevey}, {Evans}, {Eyer},
  {Hutton}, {Jansen}, {Jordi}, {Klioner}, {Lammers}, {Lindegren}, {Mignard},
  {Panem}, {Pourbaix}, {Randich}, {Sartoretti}, {Soubiran}, {Walton}, {Arenou},
  {Bailer-Jones}, {Bastian}, {Cropper}, {Drimmel}, {Katz}, {Lattanzi}, {van
  Leeuwen}, {Bakker}, {Casta{\~n}eda}, {De Angeli}, {Ducourant}, {Fouesneau},
  {Fr{\'e}mat}, {Guerra}, {Guerrier}, {Guiraud}, {Jean-Antoine Piccolo},
  {Masana}, {Messineo}, {Mowlavi}, {Nicolas}, {Nienartowicz}, {Pailler},
  {Panuzzo}, {Riclet}, {Roux}, {Seabroke}, {Sordo}, {Tanga}, {Th{\'e}venin},
  {Gracia-Abril}, {Portell}, {Teyssier}, {Altmann}, {Andrae}, {Bellas-Velidis},
  {Benson}, {Berthier}, {Blomme}, {Brugaletta}, {Burgess}, {Busso}, {Carry},
  {Cellino}, {Cheek}, {Damerdji}, {Davidson}, {Delchambre}, {Dell'Oro},
  {Fern{\'a}ndez-Hern{\'a}ndez}, {Galluccio}, {Garc{\'\i}a-Lario},
  {Garcia-Reinaldos}, {Gonz{\'a}lez-N{\'u}{\~n}ez}, {Gosset}, {Haigron},
  {Halbwachs}, {Hambly}, {Harrison}, {Hatzidimitriou}, {Heiter},
  {Hern{\'a}ndez}, {Hestroffer}, {Hodgkin}, {Holl}, {Jan{\ss}en}, {Jevardat de
  Fombelle}, {Jordan}, {Krone-Martins}, {Lanzafame}, {L{\"o}ffler}, {Lorca},
  {Manteiga}, {Marchal}, {Marrese}, {Moitinho}, {Mora}, {Muinonen}, {Osborne},
  {Pancino}, {Pauwels}, {Recio-Blanco}, {Richards}, {Riello}, {Rimoldini},
  {Robin}, {Roegiers}, {Rybizki}, {Siopis}, {Smith}, {Sozzetti}, {Ulla},
  {Utrilla}, {van Leeuwen}, {van Reeven}, {Abbas}, {Abreu Aramburu}, {Accart},
  {Aerts}, {Aguado}, {Ajaj}, {Altavilla}, {{\'A}lvarez}, {{\'A}lvarez
  Cid-Fuentes}, {Alves}, {Anderson}, {Anglada Varela}, {Antoja}, {Audard},
  {Baines}, {Baker}, {Balaguer-N{\'u}{\~n}ez}, {Balog}, {Barache}, {Barbato},
  {Barros}, {Barstow}, {Bartolom{\'e}}, {Bassilana}, {Bauchet},
  {Baudesson-Stella}, {Becciani}, {Bellazzini}, {Bernet}, {Bertone}, {Bianchi},
  {Blanco-Cuaresma}, {Boch}, {Bombrun}, {Bossini}, {Bouquillon}, {Bragaglia},
  {Bramante}, {Breedt}, {Bressan}, {Brouillet}, {Bucciarelli}, {Burlacu},
  {Busonero}, {Butkevich}, {Buzzi}, {Caffau}, {Cancelliere}, {C{\'a}novas},
  {Cantat-Gaudin}, {Carballo}, {Carlucci}, {Carnerero}, {Carrasco},
  {Casamiquela}, {Castellani}, {Castro Sampol}, {Chaoul}, {Charlot},
  {Chiavassa}, {Comoretto}, {Cooper}, {Cornez}, {Cowell}, \&
  {Crifo}}]{luri2021}
{Gaia Collaboration}, {Luri}, X., {Chemin}, L., {et~al.} 2021{\natexlab{a}},
  \aap, 649, A7

\bibitem[{{Gaia Collaboration} {et~al.}(2021{\natexlab{b}}){Gaia
  Collaboration}, {Brown}, {Vallenari}, {Prusti}, {de Bruijne}, {Babusiaux},
  {Biermann}, {Creevey}, {Evans}, {Eyer}, {Hutton}, {Jansen}, {Jordi},
  {Klioner}, {Lammers}, {Lindegren}, {Luri}, {Mignard}, {Panem}, {Pourbaix},
  {Randich}, {Sartoretti}, {Soubiran}, {Walton}, {Arenou}, {Bailer-Jones},
  {Bastian}, {Cropper}, {Drimmel}, {Katz}, {Lattanzi}, {van Leeuwen}, {Bakker},
  {Cacciari}, {Casta{\~n}eda}, {De Angeli}, {Ducourant}, {Fabricius},
  {Fouesneau}, {Fr{\'e}mat}, {Guerra}, {Guerrier}, {Guiraud}, {Jean-Antoine
  Piccolo}, {Masana}, {Messineo}, {Mowlavi}, {Nicolas}, {Nienartowicz},
  {Pailler}, {Panuzzo}, {Riclet}, {Roux}, {Seabroke}, {Sordo}, {Tanga},
  {Th{\'e}venin}, {Gracia-Abril}, {Portell}, {Teyssier}, {Altmann}, {Andrae},
  {Bellas-Velidis}, {Benson}, {Berthier}, {Blomme}, {Brugaletta}, {Burgess},
  {Busso}, {Carry}, {Cellino}, {Cheek}, {Clementini}, {Damerdji}, {Davidson},
  {Delchambre}, {Dell'Oro}, {Fern{\'a}ndez-Hern{\'a}ndez}, {Galluccio},
  {Garc{\'\i}a-Lario}, {Garcia-Reinaldos}, {Gonz{\'a}lez-N{\'u}{\~n}ez},
  {Gosset}, {Haigron}, {Halbwachs}, {Hambly}, {Harrison}, {Hatzidimitriou},
  {Heiter}, {Hern{\'a}ndez}, {Hestroffer}, {Hodgkin}, {Holl}, {Jan{\ss}en},
  {Jevardat de Fombelle}, {Jordan}, {Krone-Martins}, {Lanzafame},
  {L{\"o}ffler}, {Lorca}, {Manteiga}, {Marchal}, {Marrese}, {Moitinho}, {Mora},
  {Muinonen}, {Osborne}, {Pancino}, {Pauwels}, {Petit}, {Recio-Blanco},
  {Richards}, {Riello}, {Rimoldini}, {Robin}, {Roegiers}, {Rybizki}, {Sarro},
  {Siopis}, {Smith}, {Sozzetti}, {Ulla}, {Utrilla}, {van Leeuwen}, {van
  Reeven}, {Abbas}, {Abreu Aramburu}, {Accart}, {Aerts}, {Aguado}, {Ajaj},
  {Altavilla}, {{\'A}lvarez}, {{\'A}lvarez Cid-Fuentes}, {Alves}, {Anderson},
  {Anglada Varela}, {Antoja}, {Audard}, {Baines}, {Baker},
  {Balaguer-N{\'u}{\~n}ez}, {Balbinot}, {Balog}, {Barache}, {Barbato},
  {Barros}, {Barstow}, {Bartolom{\'e}}, {Bassilana}, {Bauchet},
  {Baudesson-Stella}, {Becciani}, {Bellazzini}, {Bernet}, {Bertone}, {Bianchi},
  {Blanco-Cuaresma}, {Boch}, {Bombrun}, {Bossini}, {Bouquillon}, {Bragaglia},
  {Bramante}, {Breedt}, {Bressan}, {Brouillet}, {Bucciarelli}, {Burlacu},
  {Busonero}, {Butkevich}, {Buzzi}, {Caffau}, {Cancelliere}, {C{\'a}novas},
  {Cantat-Gaudin}, {Carballo}, {Carlucci}, {Carnerero}, {Carrasco},
  {Casamiquela}, {Castellani}, {Castro-Ginard}, {Castro Sampol}, {Chaoul},
  {Charlot}, {Chemin}, {Chiavassa}, {Cioni}, {Comoretto}, {Cooper}, {Cornez},
  {Cowell}, {Crifo}, {Crosta}, {Crowley}, {Dafonte}, {Dapergolas}, {David},
  {David}, {de Laverny}, {De Luise}, {De March}, {De Ridder}, {de Souza}, {de
  Teodoro}, {de Torres}, {del Peloso}, {del Pozo}, {Delbo}, {Delgado},
  {Delgado}, {Delisle}, {Di Matteo}, {Diakite}, {Diener}, {Distefano},
  {Dolding}, {Eappachen}, {Edvardsson}, {Enke}, {Esquej}, {Fabre}, {Fabrizio},
  {Faigler}, {Fedorets}, {Fernique}, {Fienga}, {Figueras}, {Fouron},
  {Fragkoudi}, {Fraile}, {Franke}, {Gai}, {Garabato}, {Garcia-Gutierrez},
  {Garc{\'\i}a-Torres}, {Garofalo}, {Gavras}, {Gerlach}, {Geyer}, {Giacobbe},
  {Gilmore}, {Girona}, {Giuffrida}, {Gomel}, {Gomez}, {Gonzalez-Santamaria},
  {Gonz{\'a}lez-Vidal}, {Granvik}, {Guti{\'e}rrez-S{\'a}nchez}, {Guy},
  {Hauser}, {Haywood}, {Helmi}, {Hidalgo}, {Hilger}, {H{\l}adczuk}, {Hobbs},
  {Holland}, {Huckle}, {Jasniewicz}, {Jonker}, {Juaristi Campillo}, {Julbe},
  {Karbevska}, {Kervella}, {Khanna}, {Kochoska}, {Kontizas}, {Kordopatis},
  {Korn}, {Kostrzewa-Rutkowska}, {Kruszy{\'n}ska}, {Lambert}, {Lanza}, {Lasne},
  {Le Campion}, {Le Fustec}, {Lebreton}, {Lebzelter}, {Leccia}, {Leclerc},
  {Lecoeur-Taibi}, {Liao}, {Licata}, {Lindstr{\o}m}, {Lister}, {Livanou},
  {Lobel}, {Madrero Pardo}, {Managau}, {Mann}, {Marchant}, {Marconi}, {Marcos
  Santos}, {Marinoni}, {Marocco}, {Marshall}, {Martin Polo},
  {Mart{\'\i}n-Fleitas}, {Masip}, {Massari}, {Mastrobuono-Battisti}, {Mazeh},
  {McMillan}, {Messina}, {Michalik}, {Millar}, {Mints}, {Molina}, {Molinaro},
  {Moln{\'a}r}, {Montegriffo}, {Mor}, {Morbidelli}, {Morel}, {Morris},
  {Mulone}, {Munoz}, {Muraveva}, {Murphy}, {Musella}, {Noval}, {Ord{\'e}novic},
  {Orr{\`u}}, {Osinde}, {Pagani}, {Pagano}, {Palaversa}, {Palicio}, {Panahi},
  {Pawlak}, {Pe{\~n}alosa Esteller}, {Penttil{\"a}}, {Piersimoni}, {Pineau},
  {Plachy}, {Plum}, {Poggio}, {Poretti}, {Poujoulet}, {Pr{\v{s}}a}, {Pulone},
  {Racero}, {Ragaini}, {Rainer}, {Raiteri}, {Rambaux}, {Ramos}, {Ramos-Lerate},
  {Re Fiorentin}, {Regibo}, {Reyl{\'e}}, {Ripepi}, {Riva}, {Rixon}, {Robichon},
  {Robin}, {Roelens}, {Rohrbasser}, {Romero-G{\'o}mez}, {Rowell}, {Royer},
  {Rybicki}, {Sadowski}, {Sagrist{\`a} Sell{\'e}s}, {Sahlmann}, {Salgado},
  {Salguero}, {Samaras}, {Sanchez Gimenez}, {Sanna}, {Santove{\~n}a},
  {Sarasso}, {Schultheis}, {Sciacca}, {Segol}, {Segovia}, {S{\'e}gransan},
  {Semeux}, {Shahaf}, {Siddiqui}, {Siebert}, {Siltala}, {Slezak}, {Smart},
  {Solano}, {Solitro}, {Souami}, {Souchay}, {Spagna}, {Spoto}, {Steele},
  {Steidelm{\"u}ller}, {Stephenson}, {S{\"u}veges}, {Szabados}, {Szegedi-Elek},
  {Taris}, {Tauran}, {Taylor}, {Teixeira}, {Thuillot}, {Tonello}, {Torra},
  {Torra}, {Turon}, {Unger}, {Vaillant}, {van Dillen}, {Vanel}, {Vecchiato},
  {Viala}, {Vicente}, {Voutsinas}, {Weiler}, {Wevers}, {Wyrzykowski}, {Yoldas},
  {Yvard}, {Zhao}, {Zorec}, {Zucker}, {Zurbach}, \& {Zwitter}}]{gaia_edr3}
{Gaia Collaboration}, {Brown}, A.~G.~A., {Vallenari}, A., {et~al.}
  2021{\natexlab{b}}, \aap, 649, A1

\bibitem[{{Gonidakis} {et~al.}(2009){Gonidakis}, {Livanou}, {Kontizas},
  {Klein}, {Kontizas}, {Belcheva}, {Tsalmantza}, \&
  {Karampelas}}]{gonidakis2009}
{Gonidakis}, I., {Livanou}, E., {Kontizas}, E., {et~al.} 2009, \aap, 496, 375

\bibitem[{{Gould}(2004)}]{gould2004}
{Gould}, A. 2004, \apj, 606, 319

\bibitem[{{Gould} {et~al.}(1994){Gould}, {Miralda-Escude}, \&
  {Bahcall}}]{gould1994ApJ}
{Gould}, A., {Miralda-Escude}, J., \& {Bahcall}, J.~N. 1994, \apjl, 423, L105

\bibitem[{{Graczyk} {et~al.}(2020){Graczyk}, {Pietrzy{\'n}ski}, {Thompson},
  {Gieren}, {Zgirski}, {Villanova}, {G{\'o}rski}, {Wielg{\'o}rski},
  {Karczmarek}, {Narloch}, {Pilecki}, {Taormina}, {Smolec}, {Suchomska},
  {Gallenne}, {Nardetto}, {Storm}, {Kudritzki}, {Ka{\l}uszy{\'n}ski}, \&
  {Pych}}]{graczyk2020}
{Graczyk}, D., {Pietrzy{\'n}ski}, G., {Thompson}, I.~B., {et~al.} 2020, \apj,
  904, 13

\bibitem[{{Green}(2025)}]{green2025}
{Green}, A.~M. 2025, \jcap, 2025, 023

\bibitem[{{Green} \& {Kavanagh}(2021)}]{green2021}
{Green}, A.~M., \& {Kavanagh}, B.~J. 2021, Journal of Physics G Nuclear
  Physics, 48, 043001

\bibitem[{{Griest} {et~al.}(2013){Griest}, {Cieplak}, \& {Lehner}}]{griest2013}
{Griest}, K., {Cieplak}, A.~M., \& {Lehner}, M.~J. 2013, \prl, 111, 181302

\bibitem[{{Han} \& {Gould}(1995)}]{han1995_stat}
{Han}, C., \& {Gould}, A. 1995, \apj, 449, 521

\bibitem[{{Han} \& {Gould}(2003)}]{han_gould2003}
{Han}, C., \& {Gould}, A. 2003, \apj, 592, 172

\bibitem[{{Hawking}(1971)}]{hawking1971}
{Hawking}, S. 1971, \mnras, 152, 75

\bibitem[{{Hertzberg} {et~al.}(2021){Hertzberg}, {Nurmi}, {Schiappacasse}, \&
  {Yanagida}}]{hertzberg2021}
{Hertzberg}, M.~P., {Nurmi}, S., {Schiappacasse}, E.~D., \& {Yanagida}, T.~T.
  2021, \prd, 103, 063025

\bibitem[{{Holtzman} {et~al.}(2006){Holtzman}, {Afonso}, \&
  {Dolphin}}]{holtzman2006}
{Holtzman}, J.~A., {Afonso}, C., \& {Dolphin}, A. 2006, \apjs, 166, 534

\bibitem[{{Jacyszyn-Dobrzeniecka} {et~al.}(2016){Jacyszyn-Dobrzeniecka},
  {Skowron}, {Mr{\'o}z}, {Skowron}, {Soszy{\'n}ski}, {Udalski}, {Pietrukowicz},
  {Koz{\l}owski}, {Wyrzykowski}, {Poleski}, {Pawlak}, {Szyma{\'n}ski}, \&
  {Ulaczyk}}]{jacyszyn2016}
{Jacyszyn-Dobrzeniecka}, A.~M., {Skowron}, D.~M., {Mr{\'o}z}, P., {et~al.}
  2016, \actaa, 66, 149

\bibitem[{{Jacyszyn-Dobrzeniecka} {et~al.}(2017){Jacyszyn-Dobrzeniecka},
  {Skowron}, {Mr{\'o}z}, {Soszy{\'n}ski}, {Udalski}, {Pietrukowicz}, {Skowron},
  {Poleski}, {Koz{\l}owski}, {Wyrzykowski}, {Pawlak}, {Szyma{\'n}ski}, \&
  {Ulaczyk}}]{jacyszyn2017}
{Jacyszyn-Dobrzeniecka}, A.~M., {Skowron}, D.~M., {Mr{\'o}z}, P., {et~al.}
  2017, \actaa, 67, 1

\bibitem[{{Jiao} {et~al.}(2023){Jiao}, {Hammer}, {Wang}, {Wang}, {Amram},
  {Chemin}, \& {Yang}}]{jiao2023}
{Jiao}, Y., {Hammer}, F., {Wang}, H., {et~al.} 2023, \aap, 678, A208

\bibitem[{{Kiraga} \& {Paczy{\'n}ski}(1994)}]{kiraga1994}
{Kiraga}, M., \& {Paczy{\'n}ski}, B. 1994, \apjl, 430, L101

\bibitem[{{Koop} {et~al.}(2024){Koop}, {Antoja}, {Helmi}, {Callingham}, \&
  {Laporte}}]{koop2024}
{Koop}, O., {Antoja}, T., {Helmi}, A., {Callingham}, T.~M., \& {Laporte}, C.
  F.~P. 2024, \aap, 692, A50

\bibitem[{{Kroupa}(2001)}]{kroupa2001}
{Kroupa}, P. 2001, \mnras, 322, 231

\bibitem[{{Li} {et~al.}(2025){Li}, {Tang}, {Huang}, \& {Liu}}]{li2025}
{Li}, B., {Tang}, C.-Y., {Huang}, Z.-R., \& {Liu}, L.-H. 2025, arXiv e-prints,
  arXiv:2507.00770

\bibitem[{{Li} {et~al.}(2017){Li}, {Frenk}, {Cole}, {Wang}, \& {Gao}}]{li2017}
{Li}, R., {Frenk}, C.~S., {Cole}, S., {Wang}, Q., \& {Gao}, L. 2017, \mnras,
  468, 1426

\bibitem[{{McConnachie}(2012)}]{mcconnachie2012}
{McConnachie}, A.~W. 2012, \aj, 144, 4

\bibitem[{{Mr{\'o}z} \& {Poleski}(2018)}]{mroz_poleski2018}
{Mr{\'o}z}, P., \& {Poleski}, R. 2018, \aj, 155, 154

\bibitem[{{Mr{\'o}z} {et~al.}(2017){Mr{\'o}z}, {Udalski}, {Skowron}, {Poleski},
  {Koz{\l}owski}, {Szyma{\'n}ski}, {Soszy{\'n}ski}, {Wyrzykowski},
  {Pietrukowicz}, {Ulaczyk}, {Skowron}, \& {Pawlak}}]{mroz2017}
{Mr{\'o}z}, P., {Udalski}, A., {Skowron}, J., {et~al.} 2017, \nat, 548, 183

\bibitem[{{Mr{\'o}z} {et~al.}(2019{\natexlab{a}}){Mr{\'o}z}, {Udalski},
  {Skowron}, {Szyma{\'n}ski}, {Soszy{\'n}ski}, {Wyrzykowski}, {Pietrukowicz},
  {Koz{\l}owski}, {Poleski}, {Ulaczyk}, {Rybicki}, \& {Iwanek}}]{mroz2019b}
{Mr{\'o}z}, P., {Udalski}, A., {Skowron}, J., {et~al.} 2019{\natexlab{a}},
  \apjs, 244, 29

\bibitem[{{Mr{\'o}z} {et~al.}(2019{\natexlab{b}}){Mr{\'o}z}, {Udalski},
  {Skowron}, {Skowron}, {Soszy{\'n}ski}, {Pietrukowicz}, {Szyma{\'n}ski},
  {Poleski}, {Koz{\l}owski}, \& {Ulaczyk}}]{mroz2019c}
{Mr{\'o}z}, P., {Udalski}, A., {Skowron}, D.~M., {et~al.} 2019{\natexlab{b}},
  \apjl, 870, L10

\bibitem[{{Mr{\'o}z} {et~al.}(2024{\natexlab{a}}){Mr{\'o}z}, {Udalski},
  {Szyma{\'n}ski}, {Soszy{\'n}ski}, {Pietrukowicz}, {Koz{\l}owski}, {Poleski},
  {Skowron}, {Ulaczyk}, {Gromadzki}, {Rybicki}, {Iwanek}, {Wrona}, \&
  {Mr{\'o}z}}]{mroz2024c}
{Mr{\'o}z}, P., {Udalski}, A., {Szyma{\'n}ski}, M.~K., {et~al.}
  2024{\natexlab{a}}, \apjl, 976, L19, (Paper~III)

\bibitem[{{Mr{\'o}z} {et~al.}(2024{\natexlab{b}}){Mr{\'o}z}, {Udalski},
  {Szyma{\'n}ski}, {Kapusta}, {Soszy{\'n}ski}, {Wyrzykowski}, {Pietrukowicz},
  {Koz{\l}owski}, {Poleski}, {Skowron}, {Skowron}, {Ulaczyk}, {Gromadzki},
  {Rybicki}, {Iwanek}, {Wrona}, \& {Ratajczak}}]{mroz2024b}
{Mr{\'o}z}, P., {Udalski}, A., {Szyma{\'n}ski}, M.~K., {et~al.}
  2024{\natexlab{b}}, \apjs, 273, 4, (Paper~I)

\bibitem[{{Mr{\'o}z} {et~al.}(2024{\natexlab{c}}){Mr{\'o}z}, {Udalski},
  {Szyma{\'n}ski}, {Soszy{\'n}ski}, {Wyrzykowski}, {Pietrukowicz},
  {Koz{\l}owski}, {Poleski}, {Skowron}, {Skowron}, {Ulaczyk}, {Gromadzki},
  {Rybicki}, {Iwanek}, {Wrona}, \& {Ratajczak}}]{mroz2024a}
{Mr{\'o}z}, P., {Udalski}, A., {Szyma{\'n}ski}, M.~K., {et~al.}
  2024{\natexlab{c}}, \nat, 632, 749, (Paper~II)

\bibitem[{{Navas} {et~al.}(2024){Navas}, {Amsler}, {Gutsche}, {Hanhart},
  {Hern{\'a}ndez-Rey}, {Louren{\c{c}}o}, {Masoni}, {Mikhasenko}, {Mitchell},
  {Patrignani}, {Schwanda}, {Spanier}, {Venanzoni}, {Yuan}, {Agashe}, {Aielli},
  {Allanach}, {Alvarez-Mu{\~n}iz}, {Antonelli}, {Aschenauer}, {Asner},
  {Assamagan}, {Baer}, {Banerjee}, {Barnett}, {Baudis}, {Bauer}, {Beatty},
  {Beringer}, {Bettini}, {Biebel}, {Black}, {Blucher}, {Bonventre}, {Briere},
  {Buckley}, {Burkert}, {Bychkov}, {Cahn}, {Cao}, {Carena}, {Casarosa},
  {Ceccucci}, {Cerri}, {Chivukula}, {Cowan}, {Cranmer}, {Crede}, {Cremonesi},
  {D'Ambrosio}, {Damour}, {de Florian}, {de Gouv{\^e}a}, {DeGrand}, {Demers},
  {Demiragli}, {Dobrescu}, {D'Onofrio}, {Doser}, {Dreiner}, {Eerola}, {Egede},
  {Eidelman}, {El-Khadra}, {Ellis}, {Eno}, {Erler}, {Ezhela}, {Fava},
  {Fetscher}, {Fields}, {Freitas}, {Gallagher}, {Gershon}, {Gershtein},
  {Gherghetta}, {Gonzalez-Garcia}, {Goodman}, {Grab}, {Gritsan}, {Grojean},
  {Groom}, {Gr{\"u}newald}, {Gurtu}, {Haber}, {Hamel}, {Hashimoto}, {Hayato},
  {Hebecker}, {Heinemeyer}, {Hikasa}, {Hisano}, {H{\"o}cker}, {Holder}, {Hsu},
  {Huston}, {Hyodo}, {Ianni}, {Kado}, {Karliner}, {Katz}, {Kenzie}, {Khoze},
  {Klein}, {Krauss}, {Kreps}, {Kri{\v{z}}an}, {Krusche}, {Kwon}, {Lahav},
  {Lellouch}, {Lesgourgues}, {Liddle}, {Ligeti}, {Lin}, {Lippmann}, {Liss},
  {Lister}, {Littenberg}, {Lugovsky}, {Lugovsky}, {Lusiani}, {Makida},
  {Maltoni}, {Manohar}, {Marciano}, {Matthews}, {Mei{\ss}ner},
  {Melzer-Pellmann}, {Mertsch}, {Miller}, {Milstead}, {M{\"o}nig}, {Molaro},
  {Moortgat}, {Moskovic}, {Nagata}, {Nakamura}, {Narain}, {Nason}, {Nelles},
  {Neubert}, {Nir}, {O'Connell}, {O'Hare}, {Olive}, {Peacock}, {Pianori},
  {Pich}, {Piepke}, {Pietropaolo}, {Pomarol}, {Pordes}, {Profumo}, {Quadt},
  {Rabbertz}, {Rademacker}, {Raffelt}, {Ramsey-Musolf}, {Richardson},
  {Ringwald}, {Robinson}, {Roesler}, {Rolli}, {Romaniouk}, {Rosenberg},
  {Rosner}, {Rybka}, {Ryskin}, {Ryutin}, {Safdi}, {Sakai}, {Sarkar}, {Sauli},
  {Schneider}, {Sch{\"o}nert}, {Scholberg}, {Schwartz}, {Schwiening}, {Scott},
  {Sefkow}, {Seljak}, {Sharma}, {Sharpe}, {Shiltsev}, {Signorelli}, {Silari},
  {Simon}, {Sj{\"o}strand}, {Skands}, {Skwarnicki}, {Smoot}, {Soffer}, {Sozzi},
  {Spiering}, {Stahl}, {Sumino}, {Takahashi}, {Tanabashi}, \&
  {Tanaka}}]{navas2024}
{Navas}, S., {Amsler}, C., {Gutsche}, T., {et~al.} 2024, \prd, 110, 030001

\bibitem[{{Niikura} {et~al.}(2019){Niikura}, {Takada}, {Yasuda}, {Lupton},
  {Sumi}, {More}, {Kurita}, {Sugiyama}, {More}, {Oguri}, \&
  {Chiba}}]{niikura2019}
{Niikura}, H., {Takada}, M., {Yasuda}, N., {et~al.} 2019, Nature Astronomy, 3,
  524

\bibitem[{{Ou} {et~al.}(2024){Ou}, {Eilers}, {Necib}, \& {Frebel}}]{ou2024}
{Ou}, X., {Eilers}, A.-C., {Necib}, L., \& {Frebel}, A. 2024, \mnras, 528, 693

\bibitem[{{Paczy{\'n}ski}(1986)}]{paczynski1986}
{Paczy{\'n}ski}, B. 1986, \apj, 304, 1

\bibitem[{{Palanque-Delabrouille} {et~al.}(1998){Palanque-Delabrouille},
  {Afonso}, {Albert}, {Andersen}, {Ansari}, {Aubourg}, {Bareyre}, {Bauer},
  {Beaulieu}, {Bouquet}, {Char}, {Charlot}, {Couchot}, {Coutures}, {Derue},
  {Ferlet}, {Glicenstein}, {Goldman}, {Gould}, {Graff}, {Gros}, {Haissinski},
  {Hamilton}, {Hardin}, {de Kat}, {Lesquoy}, {Loup}, {Magneville}, {Mansoux},
  {Marquette}, {Maurice}, {Milsztajn}, {Moniez}, {Perdereau}, {Prevot},
  {Renault}, {Rich}, {Spiro}, {Vidal-Madjar}, {Vigroux}, {Zylberajch}, \& {EROS
  Collaboration}}]{palanque1998}
{Palanque-Delabrouille}, N., {Afonso}, C., {Albert}, J.~N., {et~al.} 1998,
  \aap, 332, 1

\bibitem[{Pedregosa {et~al.}(2011)Pedregosa, Varoquaux, Gramfort, Michel,
  Thirion, Grisel, Blondel, Prettenhofer, Weiss, Dubourg, Vanderplas, Passos,
  Cournapeau, Brucher, Perrot, \& Duchesnay}]{scikit-learn}
Pedregosa, F., Varoquaux, G., Gramfort, A., {et~al.} 2011, Journal of Machine
  Learning Research, 12, 2825

\bibitem[{{Pieri} {et~al.}(2011){Pieri}, {Lavalle}, {Bertone}, \&
  {Branchini}}]{pieri2011}
{Pieri}, L., {Lavalle}, J., {Bertone}, G., \& {Branchini}, E. 2011, \prd, 83,
  023518

\bibitem[{{Sahu}(1994)}]{sahu1994}
{Sahu}, K.~C. 1994, \nat, 370, 275

\bibitem[{{Sasaki} {et~al.}(2016){Sasaki}, {Suyama}, {Tanaka}, \&
  {Yokoyama}}]{sasaki2016}
{Sasaki}, M., {Suyama}, T., {Tanaka}, T., \& {Yokoyama}, S. 2016, \prl, 117,
  061101

\bibitem[{{Schlafly} \& {Finkbeiner}(2011)}]{schlafly2011}
{Schlafly}, E.~F., \& {Finkbeiner}, D.~P. 2011, \apj, 737, 103

\bibitem[{{Seng{\"u}l} {et~al.}(2022){Seng{\"u}l}, {Dvorkin}, {Ostdiek}, \&
  {Tsang}}]{sengul2022}
{Seng{\"u}l}, A.~{\c{C}}., {Dvorkin}, C., {Ostdiek}, B., \& {Tsang}, A. 2022,
  \mnras, 515, 4391

\bibitem[{{Skowron} {et~al.}(2021){Skowron}, {Skowron}, {Udalski},
  {Szyma{\'n}ski}, {Soszy{\'n}ski}, {Wyrzykowski}, {Ulaczyk}, {Poleski},
  {Koz{\l}owski}, {Pietrukowicz}, {Mr{\'o}z}, {Rybicki}, {Iwanek}, {Wrona}, \&
  {Gromadzki}}]{skowron2021}
{Skowron}, D.~M., {Skowron}, J., {Udalski}, A., {et~al.} 2021, \apjs, 252, 23

\bibitem[{{Skowron} {et~al.}(2016){Skowron}, {Udalski}, {Koz{\l}owski},
  {Szyma{\'n}ski}, {Mr{\'o}z}, {Wyrzykowski}, {Poleski}, {Pietrukowicz},
  {Ulaczyk}, {Pawlak}, \& {Soszy{\'n}ski}}]{skowron2016}
{Skowron}, J., {Udalski}, A., {Koz{\l}owski}, S., {et~al.} 2016, \actaa, 66, 1

\bibitem[{{Smith} {et~al.}(2003){Smith}, {Mao}, \& {Paczy{\'n}ski}}]{smith2003}
{Smith}, M.~C., {Mao}, S., \& {Paczy{\'n}ski}, B. 2003, \mnras, 339, 925

\bibitem[{{Springel} {et~al.}(2008){Springel}, {Wang}, {Vogelsberger},
  {Ludlow}, {Jenkins}, {Helmi}, {Navarro}, {Frenk}, \& {White}}]{springel2008}
{Springel}, V., {Wang}, J., {Vogelsberger}, M., {et~al.} 2008, \mnras, 391,
  1685

\bibitem[{{Tisserand} {et~al.}(2007){Tisserand}, {Le Guillou}, {Afonso},
  {Albert}, {Andersen}, {Ansari}, {Aubourg}, {Bareyre}, {Beaulieu}, {Charlot},
  {Coutures}, {Ferlet}, {Fouqu{\'e}}, {Glicenstein}, {Goldman}, {Gould},
  {Graff}, {Gros}, {Haissinski}, {Hamadache}, {de Kat}, {Lasserre}, {Lesquoy},
  {Loup}, {Magneville}, {Marquette}, {Maurice}, {Maury}, {Milsztajn}, {Moniez},
  {Palanque-Delabrouille}, {Perdereau}, {Rahal}, {Rich}, {Spiro},
  {Vidal-Madjar}, {Vigroux}, {Zylberajch}, \& {EROS-2
  Collaboration}}]{tisserand2007}
{Tisserand}, P., {Le Guillou}, L., {Afonso}, C., {et~al.} 2007, \aap, 469, 387

\bibitem[{{Tomaney} \& {Crotts}(1996)}]{tomaney1996}
{Tomaney}, A.~B., \& {Crotts}, A. P.~S. 1996, \aj, 112, 2872

\bibitem[{{Udalski}(2003)}]{udalski2003}
{Udalski}, A. 2003, \actaa, 53, 291

\bibitem[{{Udalski} {et~al.}(1997{\natexlab{a}}){Udalski}, {Kubiak}, \&
  {Szyma\'nski}}]{udalski1997}
{Udalski}, A., {Kubiak}, M., \& {Szyma\'nski}, M. 1997{\natexlab{a}}, \actaa,
  47, 319

\bibitem[{{Udalski} {et~al.}(1997{\natexlab{b}}){Udalski}, {Szymanski},
  {Kubiak}, {Pietrzynski}, {Wozniak}, \& {Zebrun}}]{udalski1997b}
{Udalski}, A., {Szymanski}, M., {Kubiak}, M., {et~al.} 1997{\natexlab{b}},
  \actaa, 47, 431

\bibitem[{{Udalski} {et~al.}(2015){Udalski}, {Szyma{\'n}ski}, \&
  {Szyma{\'n}ski}}]{udalski2015}
{Udalski}, A., {Szyma{\'n}ski}, M.~K., \& {Szyma{\'n}ski}, G. 2015, \actaa, 65,
  1

\bibitem[{{Vegetti} {et~al.}(2010){Vegetti}, {Koopmans}, {Bolton}, {Treu}, \&
  {Gavazzi}}]{vegetti2010}
{Vegetti}, S., {Koopmans}, L.~V.~E., {Bolton}, A., {Treu}, T., \& {Gavazzi}, R.
  2010, \mnras, 408, 1969

\bibitem[{{Vegetti} {et~al.}(2012){Vegetti}, {Lagattuta}, {McKean}, {Auger},
  {Fassnacht}, \& {Koopmans}}]{vegetti2012}
{Vegetti}, S., {Lagattuta}, D.~J., {McKean}, J.~P., {et~al.} 2012, \nat, 481,
  341

\bibitem[{{Wang} {et~al.}(2023){Wang}, {Chrob{\'a}kov{\'a}},
  {L{\'o}pez-Corredoira}, \& {Sylos Labini}}]{wang2023}
{Wang}, H.-F., {Chrob{\'a}kov{\'a}}, {\v{Z}}., {L{\'o}pez-Corredoira}, M., \&
  {Sylos Labini}, F. 2023, \apj, 942, 12

\bibitem[{{Wo{\'z}niak}(2000)}]{wozniak2000}
{Wo{\'z}niak}, P.~R. 2000, \actaa, 50, 421

\bibitem[{{Wyrzykowski} {et~al.}(2009){Wyrzykowski}, {Koz{\l}owski}, {Skowron},
  {Belokurov}, {Smith}, {Udalski}, {Szyma{\'n}ski}, {Kubiak},
  {Pietrzy{\'n}ski}, {Soszy{\'n}ski}, {Szewczyk}, \&
  {{\.Z}ebru{\'n}}}]{wyrzykowski2009}
{Wyrzykowski}, {\L}., {Koz{\l}owski}, S., {Skowron}, J., {et~al.} 2009, \mnras,
  397, 1228

\bibitem[{{Wyrzykowski} {et~al.}(2010){Wyrzykowski}, {Koz{\l}owski}, {Skowron},
  {Belokurov}, {Smith}, {Udalski}, {Szyma{\'n}ski}, {Kubiak},
  {Pietrzy{\'n}ski}, {Soszy{\'n}ski}, \& {Szewczyk}}]{wyrzyk3}
{Wyrzykowski}, {\L}., {Koz{\l}owski}, S., {Skowron}, J., {et~al.} 2010, \mnras,
  407, 189

\bibitem[{{Wyrzykowski} {et~al.}(2011{\natexlab{a}}){Wyrzykowski},
  {Koz{\l}owski}, {Skowron}, {Udalski}, {Szyma{\'n}ski}, {Kubiak},
  {Pietrzy{\'n}ski}, {Soszy{\'n}ski}, {Szewczyk}, {Ulaczyk}, \&
  {Poleski}}]{wyrzykowski2011}
{Wyrzykowski}, {\L}., {Koz{\l}owski}, S., {Skowron}, J., {et~al.}
  2011{\natexlab{a}}, \mnras, 413, 493

\bibitem[{{Wyrzykowski} {et~al.}(2011{\natexlab{b}}){Wyrzykowski}, {Skowron},
  {Koz{\l}owski}, {Udalski}, {Szyma{\'n}ski}, {Kubiak}, {Pietrzy{\'n}ski},
  {Soszy{\'n}ski}, {Szewczyk}, {Ulaczyk}, {Poleski}, \& {Tisserand}}]{wyrzyk1}
{Wyrzykowski}, {\L}., {Skowron}, J., {Koz{\l}owski}, S., {et~al.}
  2011{\natexlab{b}}, \mnras, 416, 2949

\bibitem[{{Zel'dovich} \& {Novikov}(1967)}]{zeldovich1967}
{Zel'dovich}, Y.~B., \& {Novikov}, I.~D. 1967, \sovast, 10, 602

\end{thebibliography}

\appendix
\section{SMC Proper Motion and Rotation}
\label{sec:smc_pm}

\begin{deluxetable*}{CCCCCCCCC}[t!]
\tabletypesize{\scriptsize}
\tablecaption{Parameters of the SMC Proper-Motion Model\label{tab:smc_pm}}
\tablehead{
\colhead{$\rho$} & \colhead{$\partial\mu_x/\partial x$} & \colhead{$\partial\mu_x/\partial y$} & \colhead{$\mu_{x,0}$} & \colhead{$\partial\mu_y/\partial x$} & \colhead{$\partial\mu_y/\partial y$} & \colhead{$\mu_{y,0}$} & \colhead{rms($\mu_x$)} & \colhead{rms($\mu_y$)}\\
\colhead{(deg)} & \colhead{(mas\,yr$^{-1}$\,rad$^{-1}$)} & \colhead{(mas\,yr$^{-1}$\,rad$^{-1}$)} & \colhead{(mas\,yr$^{-1}$)} & \colhead{(mas\,yr$^{-1}$\,rad$^{-1}$)} & \colhead{(mas\,yr$^{-1}$\,rad$^{-1}$)} & \colhead{(mas\,yr$^{-1}$)} & \colhead{(mas\,yr$^{-1}$)} & \colhead{(mas\,yr$^{-1}$)}}
\startdata
0 < \rho < 1 & 1.0474 &  1.3827 & 0.6737 & -1.5333 & -0.4887 & -1.2330 & 0.024 & 0.064 \\
1 < \rho < 2 & 2.3849 &  0.4965 & 0.6883 & -1.7648 &  0.0268 & -1.2429 & 0.024 & 0.075 \\
2 < \rho < 3 & 2.2517 & -0.2830 & 0.7019 & -1.8570 &  0.3592 & -1.2580 & 0.041 & 0.099 \\
3 < \rho < 4 & 1.8572 & -0.6033 & 0.7423 & -1.7579 &  0.4468 & -1.2857 & 0.069 & 0.072 \\
4 < \rho < 5 & 1.9689 & -0.5812 & 0.7527 & -1.9031 &  0.7271 & -1.2893 & 0.074 & 0.063 \\
\enddata
\end{deluxetable*}

Following~\citetalias{mroz2024a}, we used the Gaia DR3 data \citep{gaia2016,gaia_edr3} to model the kinematics of the SMC. We used a similar approach to that used by \citet{helmi2018} and \citet{luri2021}. We first selected stars brighter than $G=19.5$\,mag and located within the angular distance $\rho = 8^{\circ}$ of the center of the SMC ($\alpha_0=12.80^{\circ}$, $\delta_0=-73.15^{\circ}$; \citealt{luri2021}). We selected stars with well-measured astrometric parameters ($\mathrm{RUWE} \leq 1.4$) and parallaxes consistent with those of SMC stars ($\varpi \leq 1.0$\,mas and $\varpi/\sigma_{\varpi} < 10$), in total approximately 1.8~million objects. The positions and proper motions of all stars were transformed to Cartesian coordinates ($x$, $y$, $\mu_x$, $\mu_y$) using the orthographic projection centered on $(\alpha_0,\delta_0)$ (\citetalias{mroz2024a}). We divided the area defined by $|x|\leq 0.15$\,rad and $|y| \leq 0.15$\,rad into $100\times100$ bins, and calculated the median proper motions within each bin. We then fitted the following linear model to the binned data (using bins with at least 10 stars):
\begin{align}
\begin{split}
\mu_x &= \mu_{x,0} + \frac{\partial\mu_x}{\partial x}x + \frac{\partial\mu_x}{\partial y}y,\\
\mu_y &= \mu_{y,0} + \frac{\partial\mu_y}{\partial x}x + \frac{\partial\mu_y}{\partial y}y.
\end{split}
\end{align}
The best-fit parameters---for five annuli with a width of $1^{\circ}$---are reported in Table~\ref{tab:smc_pm}. Table~\ref{tab:smc_pm} also reports the dispersion of proper motions in each annulus.

\end{document}